\documentstyle[aps,multicol,epsf,epsfig]{revtex} 

\newcommand{\xsize}{\epsfxsize=14.0cm}

\begin{document}

\draft

\title{Beyond Blobs in Percolation Cluster Structure: The Distribution\\
of 3-Blocks at the Percolation Threshold}

\author{Gerald Paul\thanks{Electronic address: gerryp@bu.edu} and H. Eugene Stanley}

\address{Center for Polymer Studies and Department of Physics\\
Boston University, Boston, MA 02215 USA}

\date{ps.tex ~~ 7 February 2002}

\maketitle

\begin{abstract}

The incipient infinite cluster appearing at the bond percolation
threshold can be decomposed into singly-connected ``links'' and
multiply-connected ``blobs.'' Here we decompose blobs into objects known
in graph theory as 3-blocks. A 3-block is a graph that cannot be
separated into disconnected subgraphs by cutting the graph at 2 or fewer
vertices. Clusters, blobs, and 3-blocks are special cases of $k$-blocks
with $k=1$, 2, and 3, respectively. We study bond percolation clusters
at the percolation threshold on 2-dimensional square lattices and
3-dimensional cubic lattices and, using Monte-Carlo simulations,
determine the distribution of the sizes of the 3-blocks into which the
blobs are decomposed. We find that the 3-blocks have fractal dimension
$d_3=1.2\pm 0.1$ in 2D and $1.15\pm 0.1$ in 3D. These fractal dimensions are
significantly smaller than the fractal dimensions of the blobs, making
possible more efficient calculation of percolation
properties. Additionally, the closeness of the estimated values for
$d_3$ in 2D and 3D is consistent with the possibility that $d_3$ is
dimension independent. Generalizing the concept of the backbone, we
introduce the concept of a ``$k$-bone'', which is the set of all points
in a percolation system connected to $k$ disjoint terminal points (or
sets of disjoint terminal points) by $k$ disjoint paths. We argue that
the fractal dimension of a $k$-bone is equal to the fractal dimension of
$k$-blocks, allowing us to discuss the relation between the fractal
dimension of $k$-blocks and recent work on path crossing probabilities.

\end{abstract}

%\pacs{}

\begin{multicols}{2}

\section{Introduction}

Percolation is the classic model for disordered systems
\cite{Ben-Avraham00,Stauffer94,Bunde96}. For concreteness we will study
bond percolation systems in which bonds on a lattice are randomly
occupied with probability $p$. Clusters are defined as groups
of sites and bonds which are connected by occupied bonds. Clusters can
be decomposed into objects known as blobs. Blobs are sets of sites and
bonds which cannot be decomposed into disconnected sets by cutting only
one bond. Equivalently blobs are sometimes described as being
multiply-connected---there are at least two disjoint paths between each
point in a blob and every other point in the blob. The decomposition of
the entire percolation cluster into blobs has been extensively studied
\cite{Gyure95}, as has the distribution of sizes of blobs in the
backbone \cite{Herrmann84}. For both cluster and backbone blobs, the
fractal dimension of the blobs is the fractal dimension of the backbone.

Here we address the questions of (i) whether there are more fundamental
objects into which blobs can be decomposed, and (ii) whether these
objects then be further decomposed.  To answer these questions, we
employ the language of graph theory, in which sites are the vertices and
bonds are the edges of a graph \cite{Tutte84}.  

\begin{figure}
\centerline{
\epsfxsize=19.0cm
\epsfclipon
\epsfbox{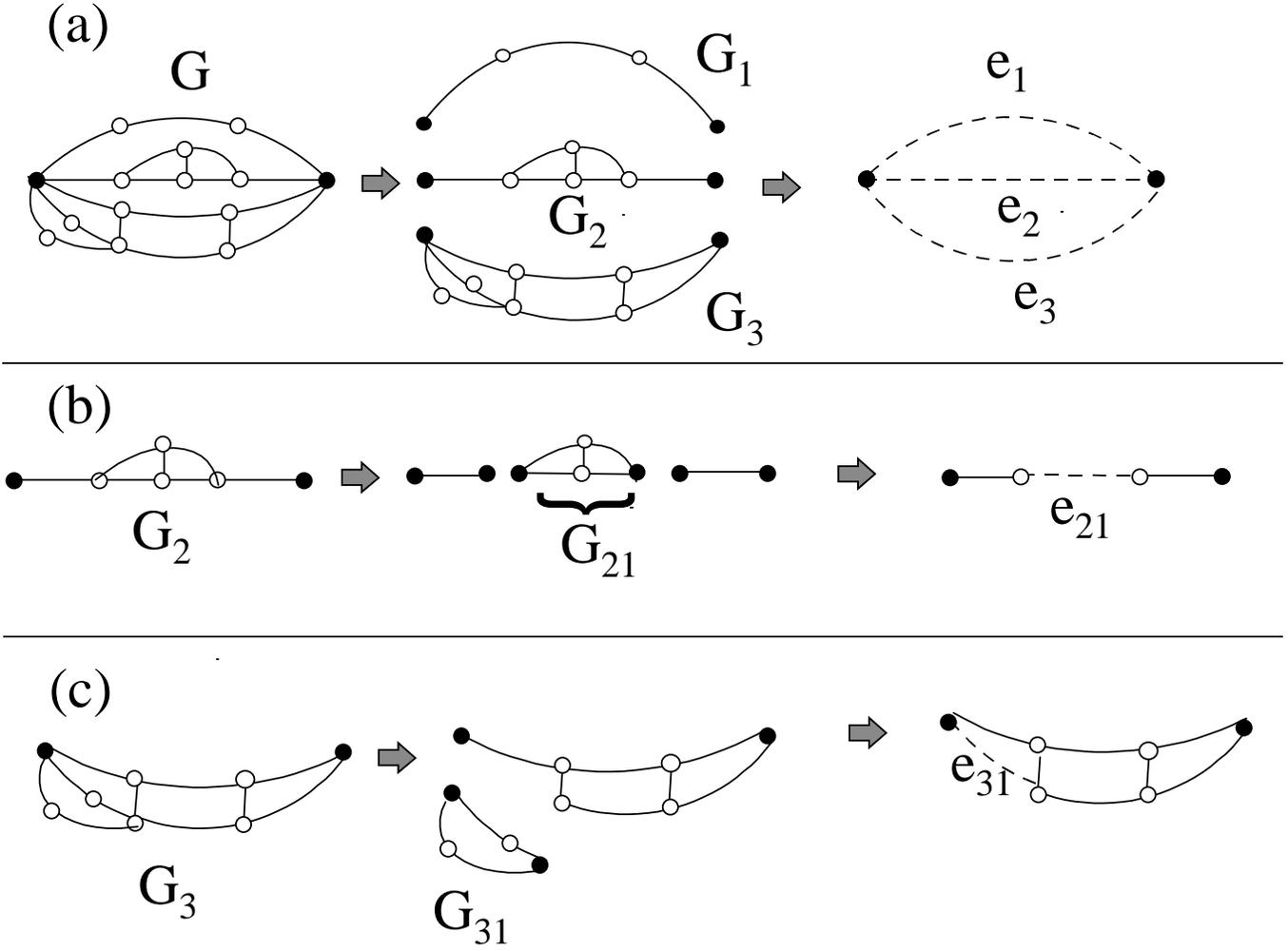}
}
\caption{(a) Decomposition of 2-block $G$ into subgraphs $G_1$, $G_2$,
and $G_3$. The rightmost graph represents $G$ with the sub-graphs
replaced by equivalent ``virtual edges.'' (b) Subgraph $G_2$ of $G$ is
decomposed by identifying subgraph $G_{21}$. The rightmost graph
represents $G_2$ with the subgraph $G_{21}$ replaced by it equivalent
edge. (c) Subgraph $G_3$ of $G$ is decomposed by identifying subgraph
$G_{31}$. The rightmost graph represents $G_3$ with the subgraph
$G_{31}$ replaced by it equivalent edge. In (a), (b), and (c) virtual
edges are denoted by dashed lines. Note that while not shown in this
figure, subgraph $G_{31}$ could be further decomposed. The 3-blocks
contained in the graph $G$ are $G_{21}$, having 5 edges, and $G_3$
(with the subgraph $G_{31}$ replaced by its equivalent edge) having 8
edges.}
\label{tpoDecomp}
\end{figure}

\begin{figure}
\centerline{
\xsize
\epsfclipon
\epsfbox{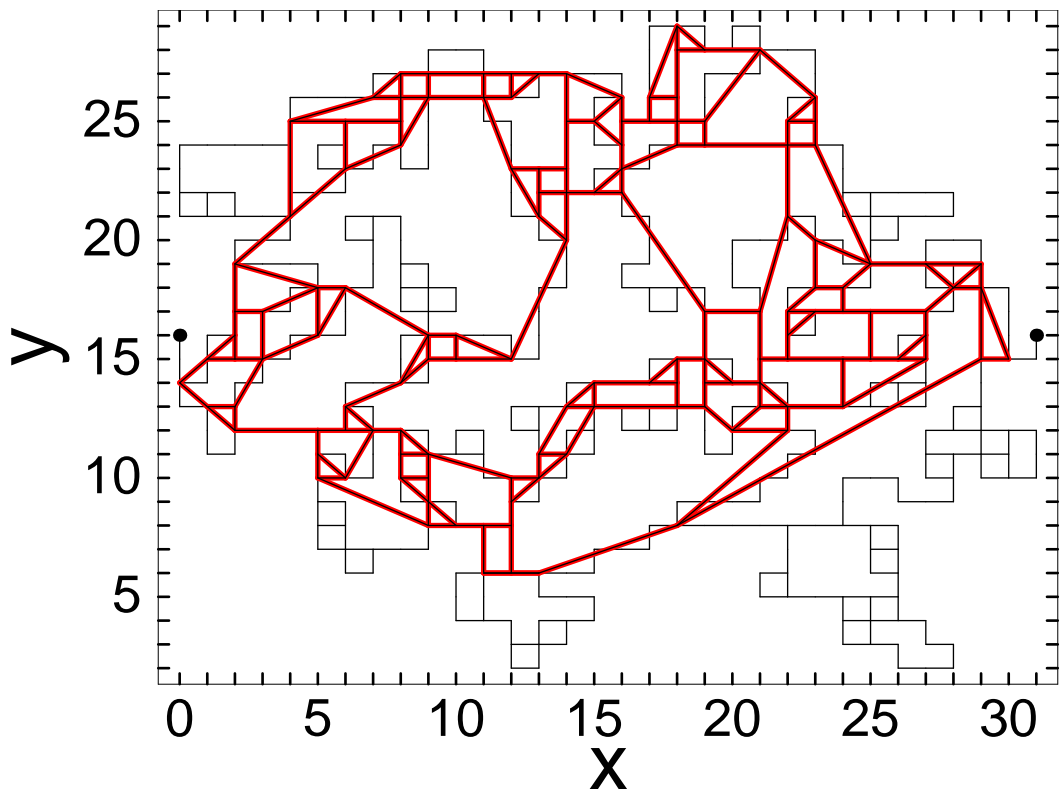}
}

\caption{Example of decomposition of backbone into 3-blocks. The thin
lines represent the bonds in the backbone between points \{0,15\} and
\{31,15\} on a lattice with L=32.  The backbone is composed of a few
single bond blobs connected to the terminal points and a single large
blob containing 950 bonds.  The thick lines represent the virtual bonds
of a single top-level 3-block into which the blob has been decomposed.
This 3-block contains 216 virtual bonds.  Some of the groups of bonds
replaced by virtual bonds can themselves be decomposed into lower level
3-blocks and so on.}
\label{preal}
\end{figure}

One can define $k$-connected graphs (or $k$-blocks) as graphs which
cannot be separated into disconnected subgraphs by cutting the graph at
fewer than $k$ vertices \cite{Tutte84,Text1}. Thus, clusters are
1-blocks and blobs are 2-blocks.  The natural next level of
decomposition of percolation systems is to decompose blobs(2-blocks)
into 3-blocks. By the definition above, 3-blocks are graphs which cannot
be decomposed by cutting the graphs at fewer than 3 vertices. From a
physicist's point of view, one can understand what 3-blocks are by
considering a blob as a resistor network with each bond being a
resistor. Assume one is trying to determine the resistance between 2
vertices of the network. One can simplify the network by using
Kirchoff's Laws to replace groups of sequential bonds and groups of
parallel bonds by single virtual bonds having resistance equivalent to
the bonds replaced. After this has been done as completely as possible,
what are left are 3-blocks. We define the mass of a 3-block as the
number of virtual bonds plus the number of non-replaced original bonds
remaining in the 3-block. Figures \ref{tpoDecomp} and \ref{preal} 
provide examples of the decomposition of a blob into 3-blocks.  It
has been shown \cite{Tutte84} that the decomposition of 2-blocks into
3-blocks is unique.

Determining the scaling of the distribution of the 3-blocks into which
the 2-blocks can be decomposed is the subject of this paper. In graph
theory, the sites are typically not constrained to a lattice structure,
and one is only concerned with the topology of the graphs; we will,
however, work on square and cubic lattices.

\section{Notation}

Because we deal with a number of different types of fractal objects, we
employ the following notation:

\begin{itemize}

\item[{(i)}] The fractal dimension of an object of type $X$ will be
denoted as $d_X$.

\item[{(ii)}] The number distribution of objects of type $X$ in space
$Y$ of size $L$ will be denoted as $n(N_X,L_Y)$.

\item[{(iii)}] The exponent of the power-law regime of a distribution of 
objects of type $X$ in space of type $Y$ will be denoted as
$\tau_{X,Y}$.

\item[{(iv)}] The amplitude of a distribution of objects of type $X$ in
space of type $Y$ will be denoted as $A_{X,Y}$.

\item[{(v)}] We define $d_{nY}$ through the relation

\begin{equation}
\label{ex10}
\langle n(L)\rangle\sim L^{d_{nY}},
\end{equation}
where $\langle n(L)\rangle$ is the average number of disjoint objects of
a given type in space $Y$.

\item[{(vi)}] We use $0,1,2,3\ldots$ to denote $k$-blocks with
$k=0,1,2,3\ldots$ corresponding to Euclidean space, clusters, blobs, and
3-blocks respectively. We use $B$ to denote the percolation cluster
backbone.

\item[{(vii)}] Additionally, because, as noted above, objects such as
3-blocks can be nested, we denote quantities that relate to all levels
of nesting with an asterisk. Specifically, $\tau_{X,Y}^\ast$ and
$A_{X,Y}^\ast$ denote the exponent of the power-law regime and the
amplitude of a distribution of nested objects of type $X$ at all levels
of nesting in space of type $Y$. Similarly, $d_{nY}^\ast$ is defined
through the relation
\begin{equation}
\label{ex10a}
\langle n^\ast(L)\rangle\sim L^{d_{nY}^\ast},
\end{equation}
where $\langle n^\ast(L)\rangle$ is the average number of nested objects
at all levels of nesting of a given type in space $Y$. Quantities not
qualified with an asterisk will denote quantities at a single level or
quantities which cannot be nested.

\end{itemize}

Using this notation, previous results are \cite{Herrmann84}
\begin{equation}
\label{ex20}
n(N_2,L_B)=A_{2,B}L^{d_{nB}}N_2^{-\tau_{2,B}}f_L\left({N_2\over
L^{d_2}}\right)
\end{equation}
for the number distribution of blobs of mass $N_2$ in the percolation
cluster backbone and \cite{Gyure95}
\begin{equation}
\label{ex21}
n(N_2,L_1)=A_{2,1}L^{d_{n1}}N_2^{-\tau_{2,1}}f_L\left({N_2\over
L^{d_2}}\right)
\end{equation}
for the number distribution of blobs of mass $N_2$ in the whole
percolation cluster. The finite-size scaling function $f_L(x)$ in
Eqs.~(\ref{ex20}) and (\ref{ex21}) approaches 0 when $x>1$ and is 1
otherwise. 

In analogy with Eqs.~(\ref{ex20}) and (\ref{ex21}) we expect the number
distribution of 3-blocks at all levels of nesting in a blob to be
\begin{equation}
n^\ast(N_3,L_2)=A_{3,2}^\ast L^{d_{n2}^\ast}N_3^{-\tau_{3,2}^\ast}f_c
\left({N_3\over c}\right)f_L\left({N_3\over L^{d_2}}\right), 
\label{e1}
\end{equation}
where $c$ is the mass of the smallest 3-block and the finite-size
scaling function $f_c(x)$ approaches 0 when $x<1$ and is 1 otherwise,
reflecting the fact that there cannot be any 3-blocks smaller than the
smallest size $c$. In all dimensions and for all lattices, $c=5$. For
simplicity we will approximate $n^\ast(N_3,L_B)$ as
\begin{equation}
n^\ast(N_3,L_2)=\cases{
A_{3,2}^\ast L^{d_{n2}^\ast}N_3^{-\tau_{3,2}^\ast} & $c\leq N_3\leq aL^{d_2}$\cr
                  0 & otherwise.}
\label{e2}
\end{equation}

\section{Simulations}

We perform simulations with $p=0.5$, the exact percolation threshold
for 2D \cite{Stauffer94,Bunde96} and $p=0.2488126$, the most precise
current estimate for the percolation threshold for 3D
\cite{LorenzXX}. We created percolation clusters which included the
sites $(0,L/2)$ and $(L,L/2)$ for the 2D simulations and the sites
$(0,L/2,L/2)$ and $(L,L/2,L/2)$ for the 3D simulations, decomposed the
backbones determined by these sites into blobs and then decomposed the
blobs into 3-blocks. We study both distributions of 3-blocks in blobs of
given mass, $N_2$, and distributions of 3-blocks in backbones in systems
of a given size, $L$.  For purposes of analysis, we group together blobs
with mass $2^{m-1}<N_2<2^m$.

We perform the decomposition into 3-blocks along the lines of the
procedure sketched in Ref.~\cite{Tutte84}. Basically, this procedure is
as follows: We first designate the blob that we are decomposing as the
2-block graph $G$. The natural next level of decomposition is to
identify connected subgraphs with two or more edges that are connected
to $G$ at only two vertices.  We denote these subgraphs
$G_1,G_2,G_3,\ldots$ of $G$ as 2-terminal objects. These 2-terminal
objects can then be replaced in $G$ by ``virtual edges,''
$e_1,e_2,e_3,\ldots$ Note that this process can be continued
recursively. That is, the subgraph $G_i$ may itself contain sub-graphs,
$G_{i1},G_{i2},G_{i3},\ldots$ that are connected to $G_i$ at only two
vertices; we then replace the subgraphs $G_{ij}$ in $G_i$ by virtual
edges $e_{Gij}$. The process continues until the only remaining
subgraphs are those that cannot be decomposed further by making cuts at
two vertices; these, by definition, are 3-blocks. An example of this
decomposition is shown in Fig.~\ref{tpoDecomp}. Other methods of
decompostion into 3-blocks are described in
Refs.~\cite{Hohberg92,Hopcroft73}.

We perform at least 3700 realizations for each system size; for the
smaller system sizes for which the simulations run more quickly we
performed as many as $10^8$ realizations. Because, the larger the
systems the larger the number of 3-blocks contained in the system, the
statistics for the larger systems was acceptable despite the lower
number of realizations. We bin the results for all system sizes in
order to smooth the plots.

\section{Two Spatial Dimensions}

In this section we discuss our results for 3-blocks in 2D
percolation. Results in 3D are analogous and are discussed in the next
section.

\subsection{3-blocks in Blobs}

\begin{figure}
\centerline{
\xsize
\epsfclipon
\epsfbox{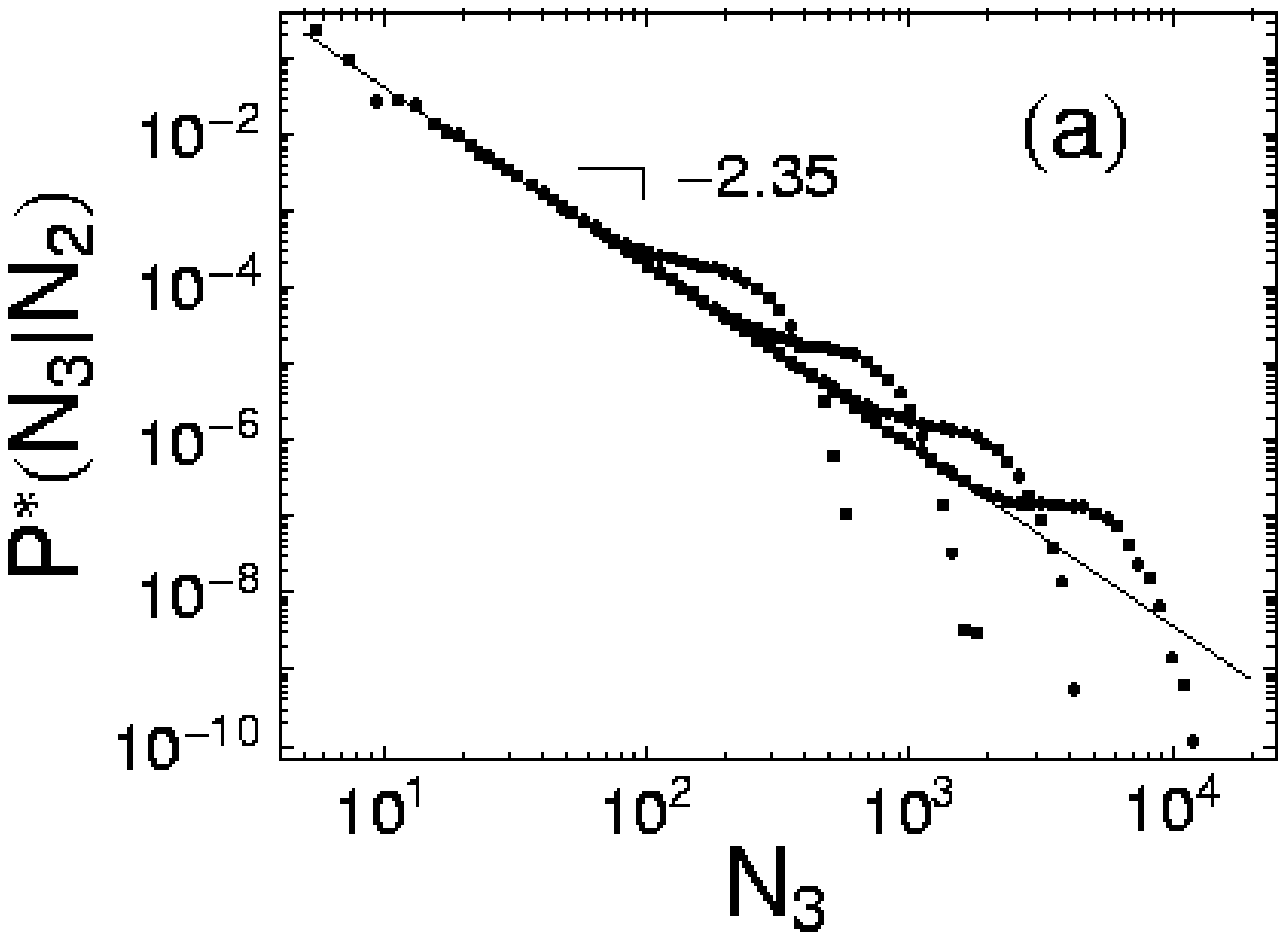}
}
\centerline{
\xsize
\epsfclipon
\epsfbox{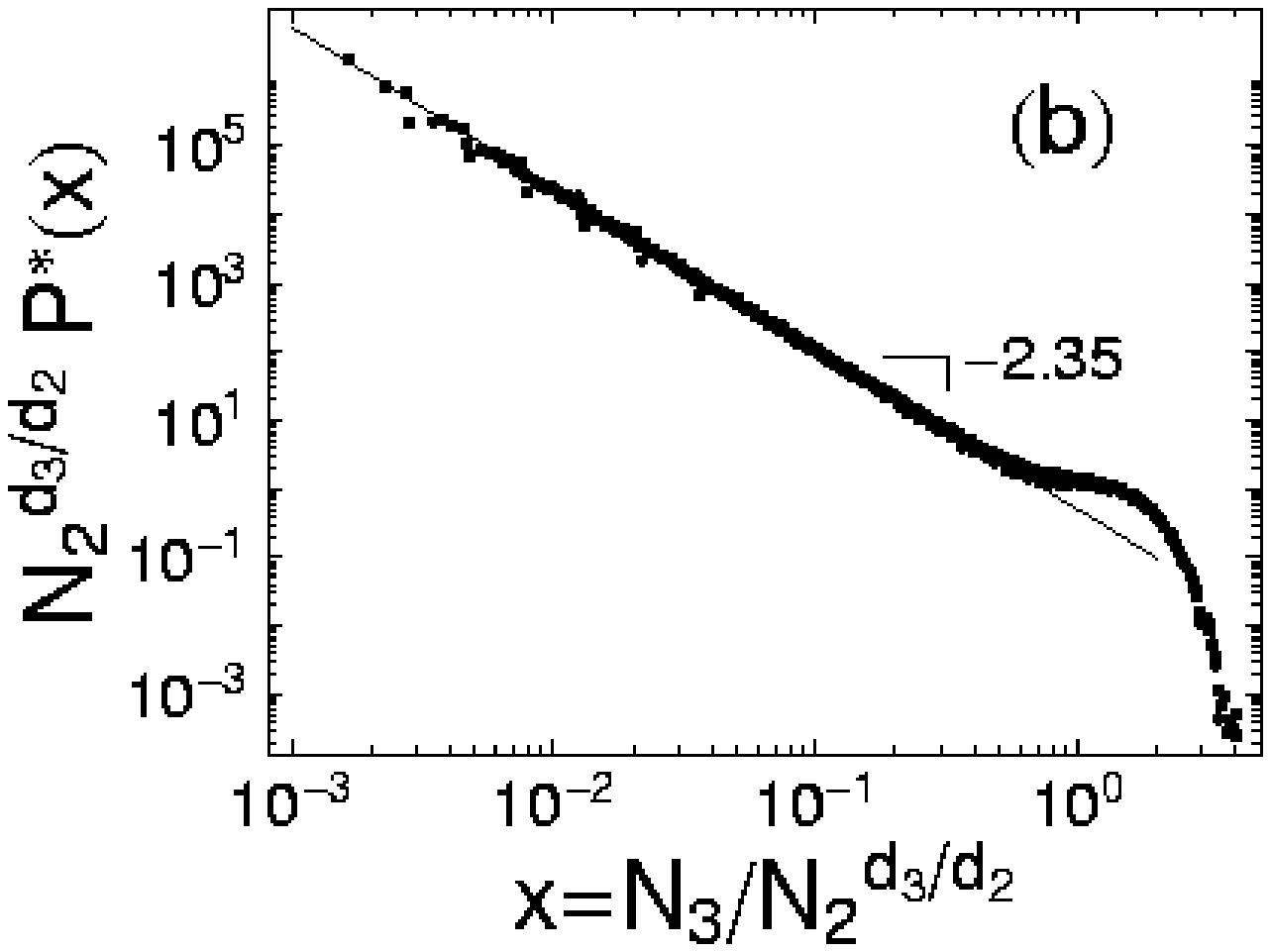}
}
\caption{2D (a) Distributions $P^\ast(N_3|N_2)$ of the number of 3-blocks of
mass $N_3$ in a blob of size $N_2$ versus $N_3$ for (from bottom to top)
$N_2=2^{10}, 2^{12}, 2^{14}$, and $2^{16}$. The distributions exhibit a power-law
regime with slope $-2.35\pm 0.05$ (b) Distributions for $N_2=2^{12}, 2^{13},
2^{14}, 2^{15}$, and $2^{16}$ scaled with the value 1.20 for the fractal
dimension $d_3$ which gives the best collapse of the plots in (a).}
\label{pAll3}
\end{figure}

Figure~\ref{pAll3}(a) plots the distributions $P^\ast(N_3|N_2)$, the
probability that a 3-block contained in a blob of size $N_2$ contains
$N_3$ bonds, for various values of $N_2$.  $P(N_3|N_2)$ is the number
distribution $n^\ast(N_3,N_2)$ normalized to unity.  Consistent with
Eqs.~(\ref{e1}) and (\ref{e2}), the plots exhibit power-law regimes
followed by cut-offs due to the finite size of the blobs.  The ``bumps''
in the distributions right before the cutoffs represent 3-blocks which
would have been larger but are truncated due to the finite size of the
blobs in which they are embedded. We estimate the slope of the power law
regimes, $\tau_{3,2}^\ast$, to be $2.35\pm 0.05$.  Since
\begin{equation}
N_3\sim L^{d_3}
\label{e101}
\end{equation}
and
\begin{equation}
N_2\sim L^{d_2}
\label{e102}
\end{equation}
we expect
\begin{equation}
N_3\sim {N_2}^{d_3/d_2}.
\label{e103}
\end{equation}

In Fig.~\ref{pAll3}(b), we show the collapsed plots in which we scale
the distributions by ${N_2}^{d_3/d_2}$ using the most precise published
estimate for $d_2$, $1.6432\pm 0.0008$ \cite{Grassberger99}. (A
consistent more recent estimate, $d_2=1.6431\pm 0.0006$, is given in
Ref.~\cite{JacobsenXX}.)  Visually, we find the best collapse is
obtained for $d_3=1.20\pm 0.1$.

We can also estimate $d_3$ using Eq.~(\ref{e220}) from the appendix
\begin{equation}
d_{\mbox{\scriptsize 3}}(\tau_{\mbox{\scriptsize
3,2}}^\ast-1)=d_{n2}^\ast=d_2.
\label{e104}
\end{equation}

Using $\tau_{3,2}^\ast=2.35\pm 0.05$ and $d_2=1.6432\pm 0.0008$, results in
an estimate of $d_3=1.22\pm 0.05$.

\subsection{3-blocks in Backbone}

\begin{figure}
\centerline{
\xsize
\epsfclipon
\epsfbox{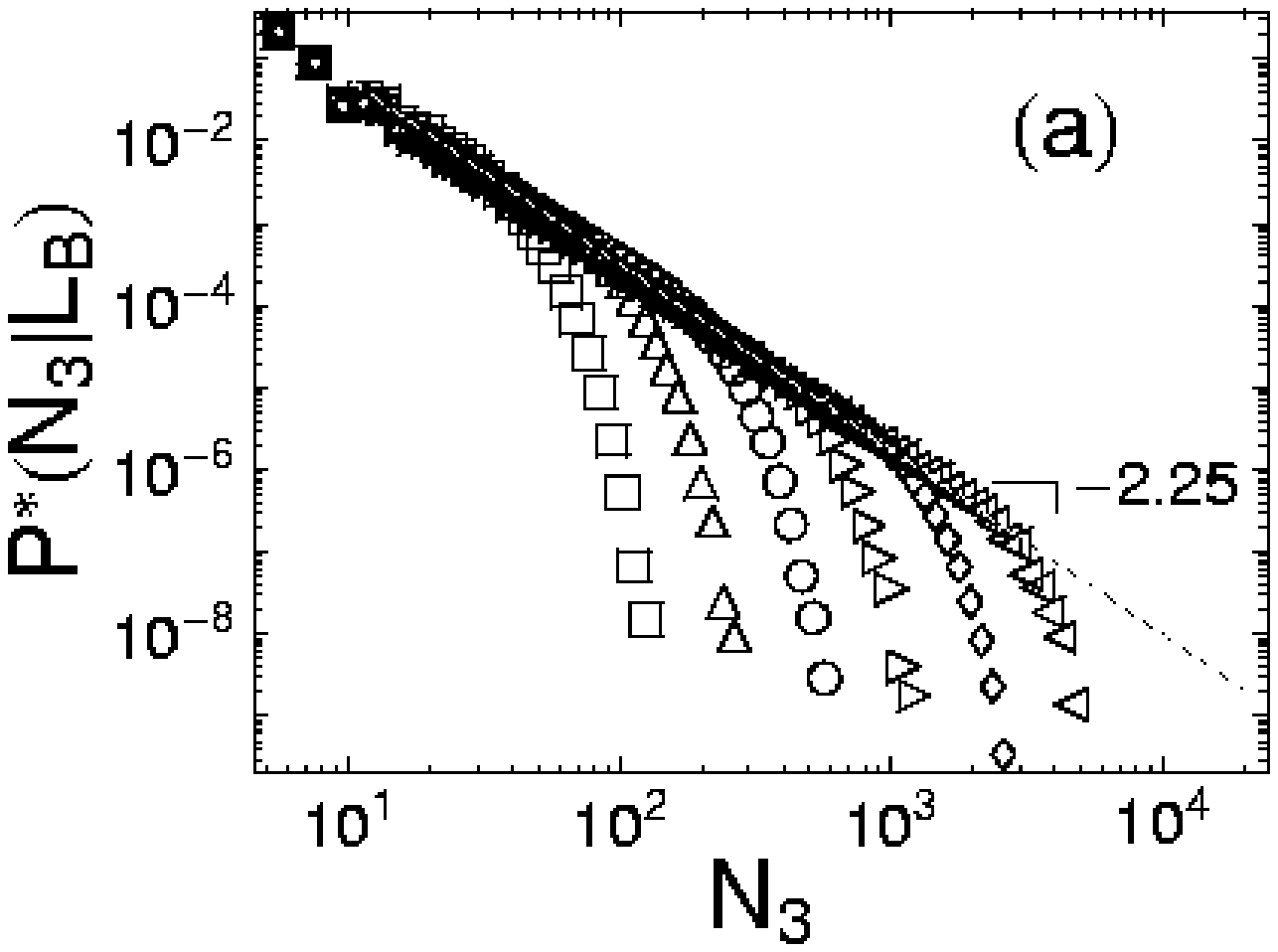}
}
\centerline{
\xsize
\epsfclipon
\epsfbox{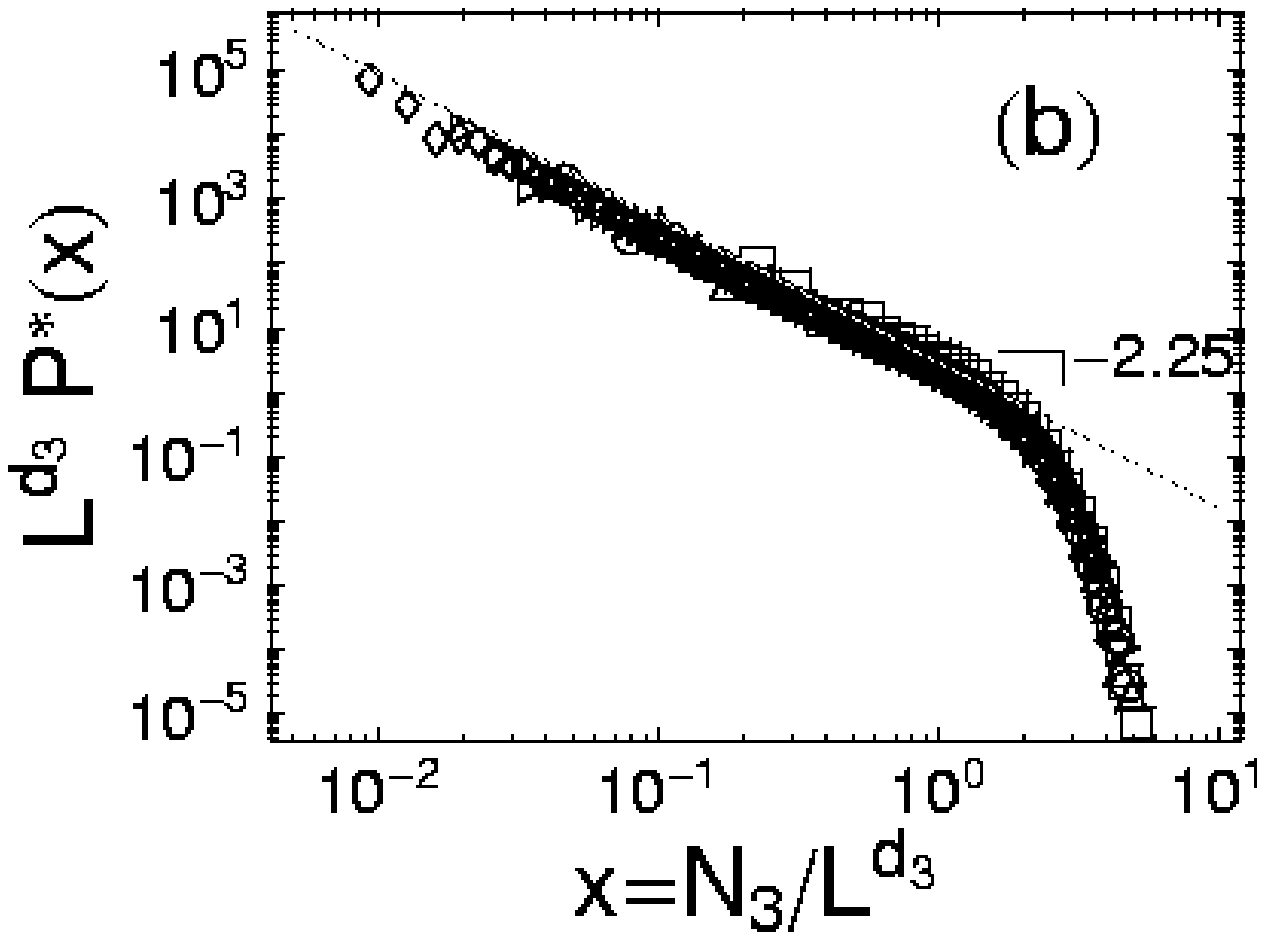}
}
\caption{2D (a) Distributions $P^\ast(N_3|L)$ of the number of 3-blocks of mass
$N_3$ in a backbone of size $L$ versus $N_3$ for (from bottom to top)
$L=16$, 32, 64, 128, 256 and 512. The distributions exhibit a power-law
regime with slope $-2.25\pm 0.05$ (b) Distributions scaled with the
value 1.15 for the fractal dimension $d_3$ which gives the best collapse
of the plots in (a).}
\label{pc}
\end{figure}

Figure~\ref{pc}(a) plots the distributions $P^\ast(N_3|L_B)$, the
probability that a 3-block contained in the backbone of a system of size
$L$ contains $N_3$ bonds, for various values of $L$. $P^\ast(N_3|L_B)$
is the number distribution $n^\ast(N_3,L_B)$ normalized to unity.  Consistent
with Eqs.~(\ref{e1}) and (\ref{e2}), the plots exhibit power-law regimes
followed by cut-offs due to the finite size of the systems. We estimate
the slope of the power law regimes, $\tau_{3,B}^\ast$, to be $2.25\pm
0.05$. In Fig.~\ref{pc}(b), we show the collapsed plots in which we scale
the distributions by $L^{d_3}$. Visually, we find the best collapse is
obtained for $d_3=1.15\pm 0.1$.

\begin{figure}
\centerline{
\xsize
\epsfclipon
\epsfbox{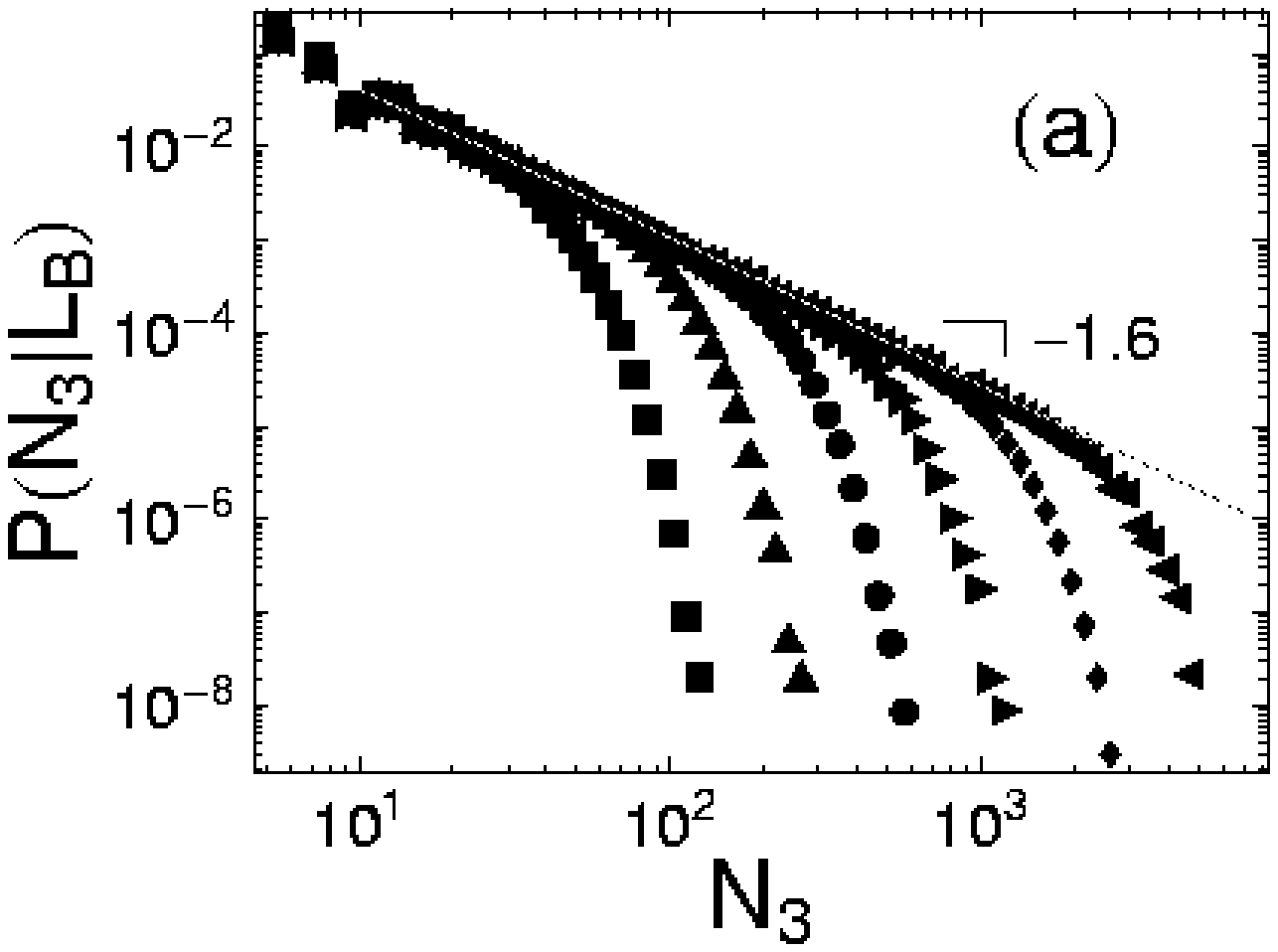}
}
\centerline{
\xsize
\epsfclipon
\epsfbox{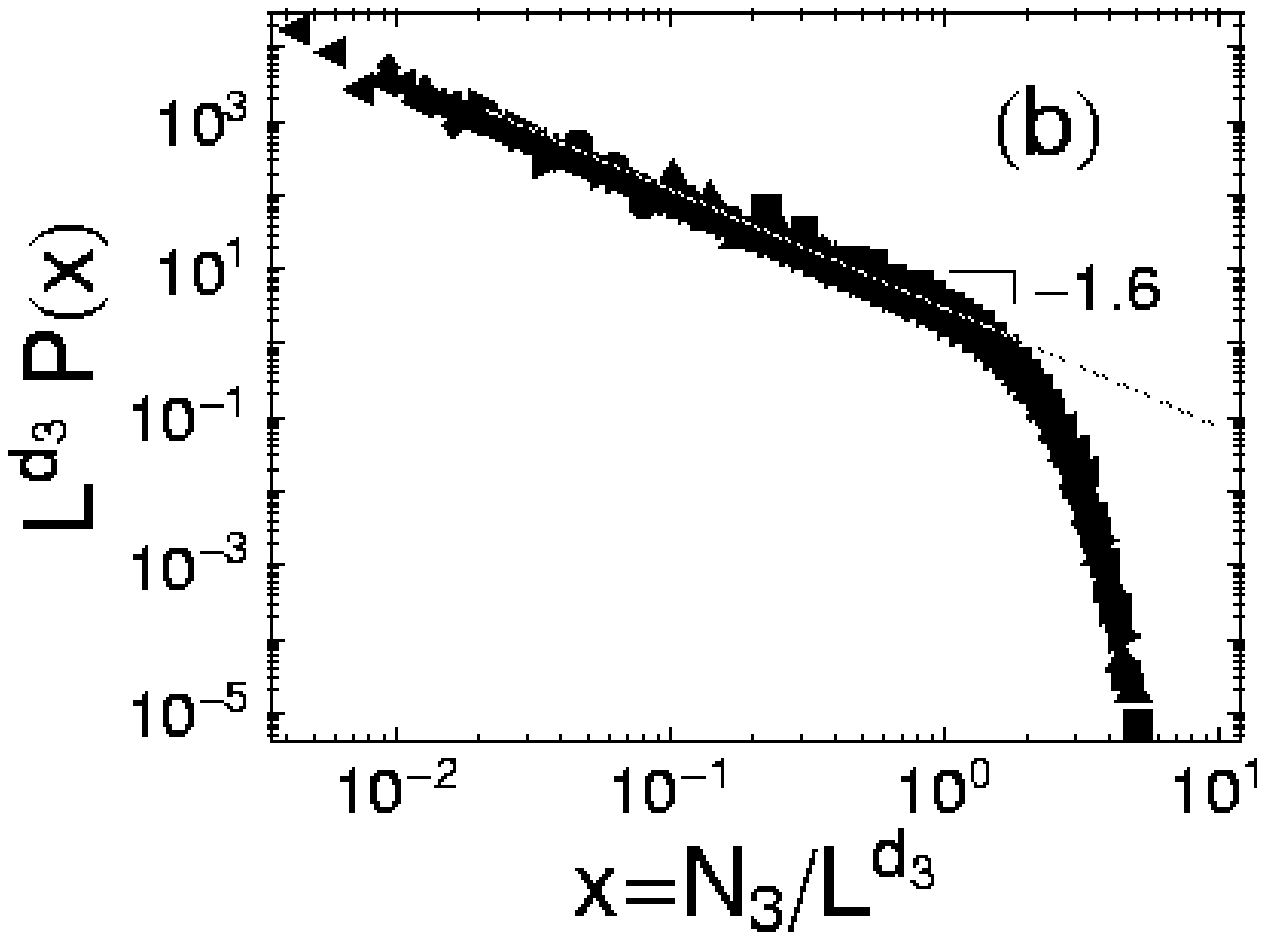}
}
\caption{2D (a) Distributions $P(N_3|L)$ of the number of top level
3-blocks of mass $N_3$ in a backbone of size $L$ versus $N_3$ for (from
top to bottom) $L=8$, 16, 32, 64, and 128. The distributions exhibit a
power-law regime with slope $-1.6\pm 0.1$. (b) Distributions scaled with
the value 1.15 for the fractal dimension $d_3$ which gives the best
collapse of the plots in (a).}
\label{pc1}
\end{figure}

\begin{figure}
\centerline{
\xsize
\epsfclipon
\epsfbox{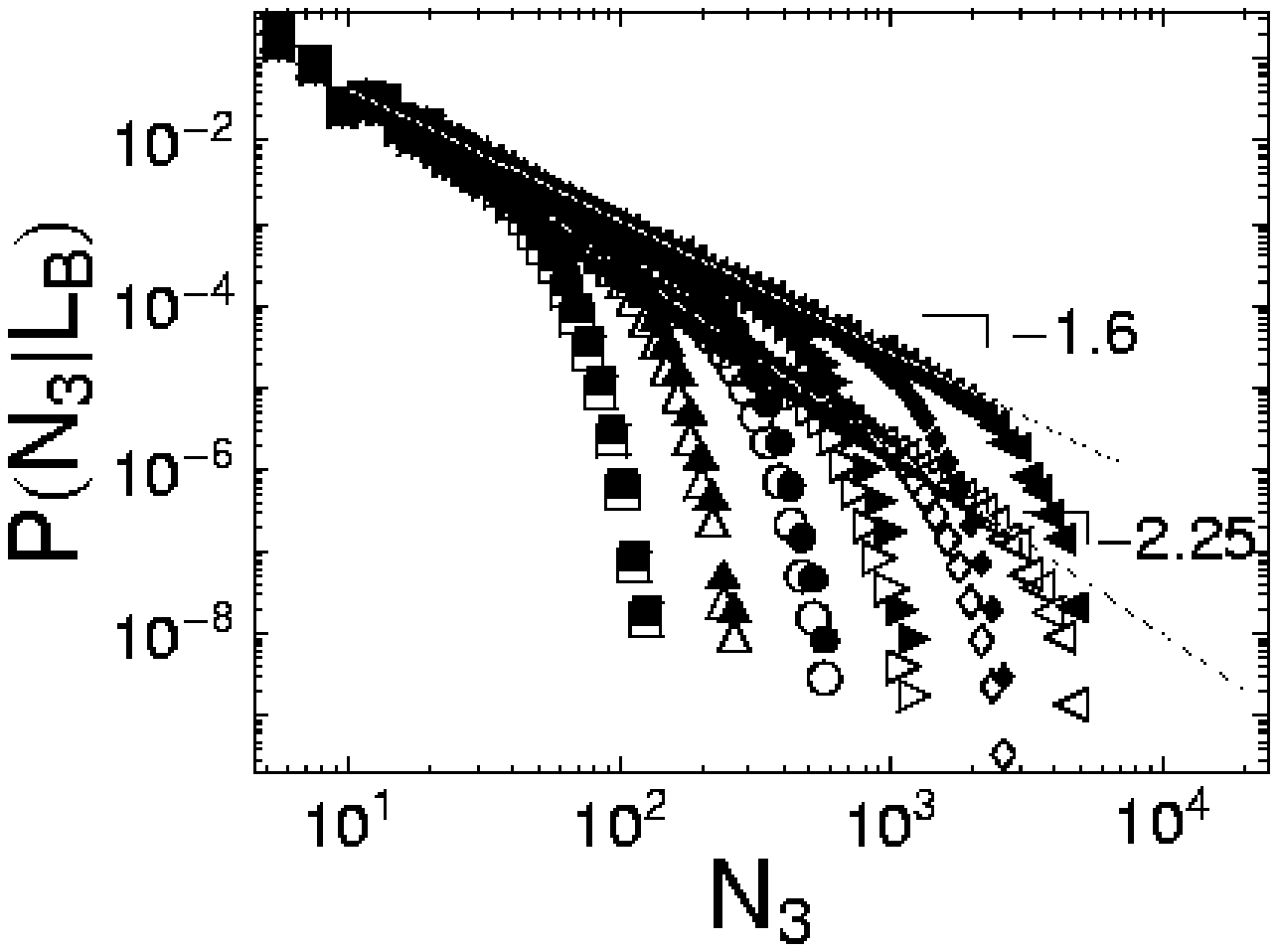}
}
\caption{2D Distributions $P(N_3|L)$ of top level 3-blocks (filled
symbols) and $P^\ast(N_3|L)$ of all-level 3-blocks (unfilled symbols). While
the slopes of the power law regimes of the two types of distributions
are different, the finite-size-system cutoffs are essentially
superimposed, consistent with the fractal dimension of the two types of
distributions being equal.}
\label{pcComb1}
\end{figure}

Next we consider the distribution of ``top-level'' 3-blocks in the
backbone.  Top-level 3-blocks are those not contained within another
3-block. In Fig.~\ref{pc1}(a), we plot the distributions $P(N_3|L_B)$, the
probability that a top-level 3-block contained in the backbone of a
system of size $L$ contains $N_3$ bonds, for various values of $L$. The
plots exhibit power-law regimes followed by cut-offs due to the finite
size of the systems.  The exponent of the power-law regimes $\tau_{3,B}$
is estimated to be $1.6\pm 0.05$.  In Fig.~\ref{pc1}(b), we show the
collapsed plots, in which we scale the distributions by $L^{d_3}$. The
best collapse is obtained for $d_3=1.15\pm 0.1$, the same value as for
the distributions of 3-blocks of all levels. Thus the fractal dimensions
of the top level 3-blocks is the same as the fractal dimension of
3-blocks of all levels but the slopes of the power law regimes are
different; this is seen also in Fig.~\ref{pcComb1}. 

We can also use Eq.~(\ref{e210})
\begin{equation}
d_{\mbox{\scriptsize 3}}(\tau_{\mbox{\scriptsize
3,B}}-1)=d_{nB}={1\over\nu}.
\label{e106}
\end{equation}
to obtain an estimate of $d_3$. Since $d_{nB}$ is known exactly in 2
dimensions and has been well studied in higher dimensions and because
one can usually determine the slope $\tau_{3,B}$ more accurately than
$d_3$ can be determined by finding the best scaling collapse, we
determine $d_3$ more accurately by solving Eq.~(\ref{e106}) for
$d_3$. Using our estimate for $\tau_{3,B}$ above we find $d_3=1.25\pm
0.1$. Combining this result with our earlier estimates, we make the final
estimate
\begin{equation}
d_3=1.20\pm 0.1.
\label{e900}
\end{equation}

\subsection{Why the Fractal Dimension of 3-blocks is Smaller than the
Fractal Dimension of the Backbone and 2-blocks}

\begin{figure}
\centerline{
\xsize
\epsfclipon
\epsfbox{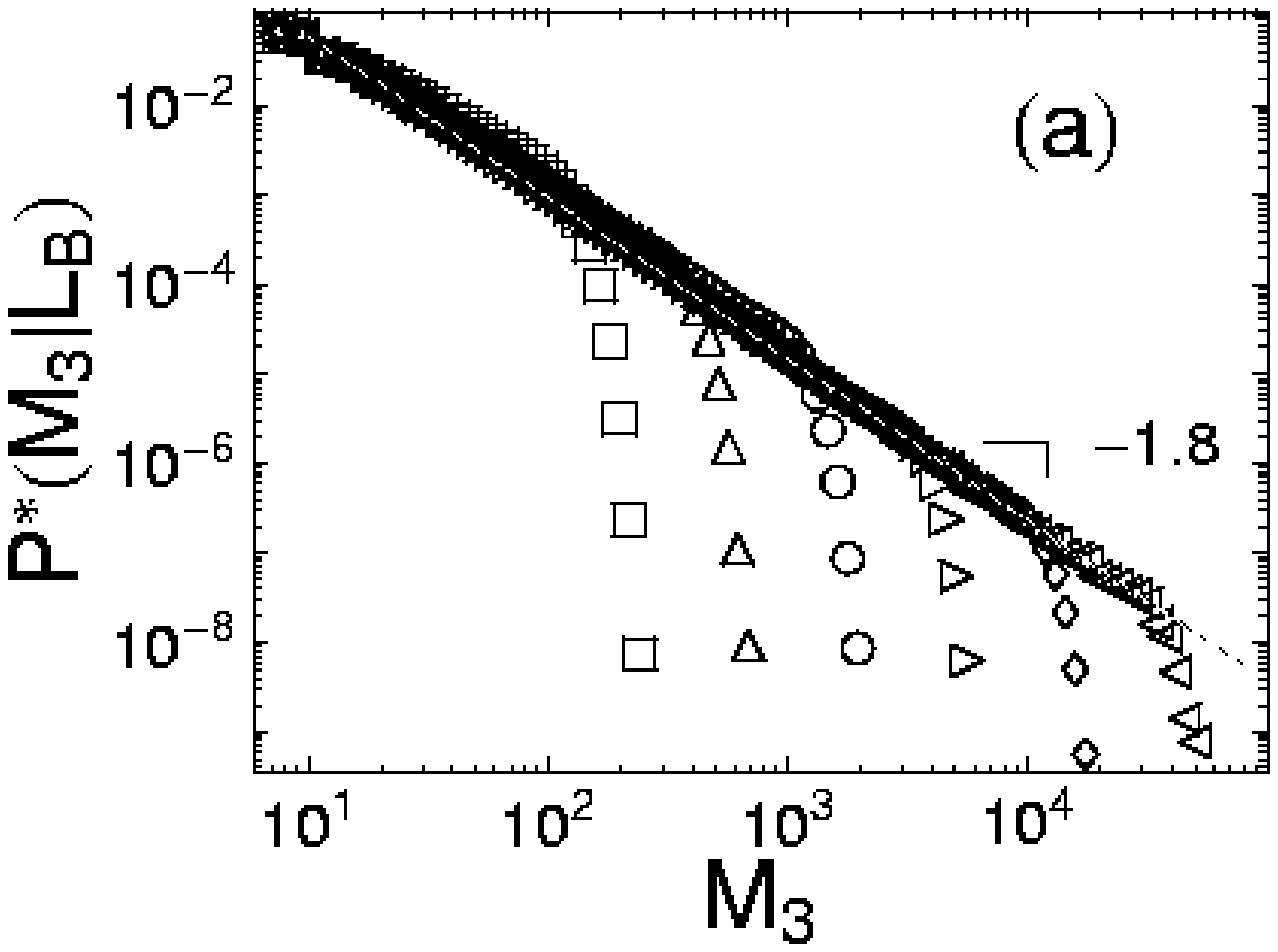}
}
\centerline{
\xsize
\epsfclipon
\epsfbox{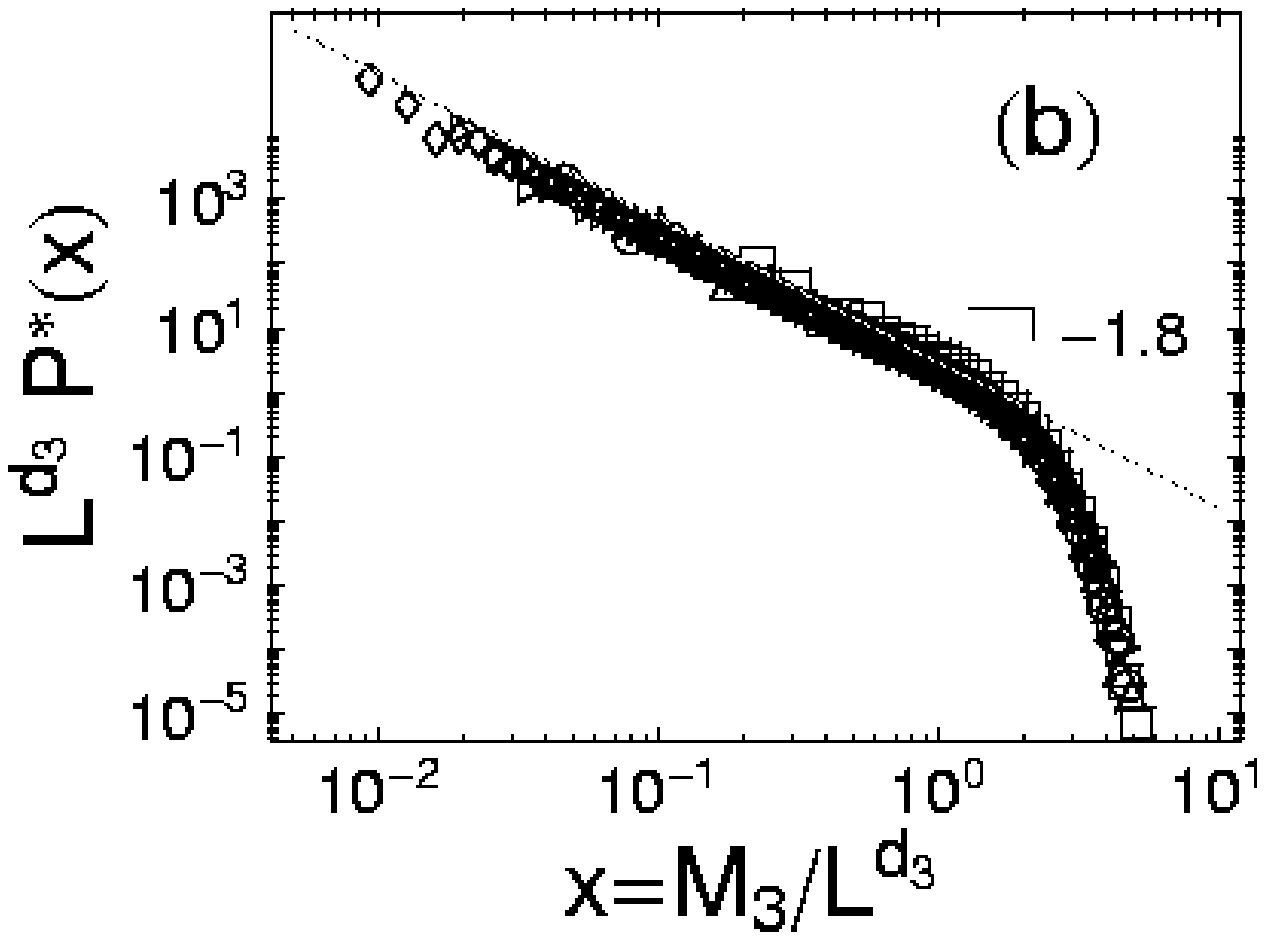}
}
\caption{2D (a) Distributions $P^\ast(M_3|L)$ of the number of 3-blocks of
mass $M_3$ in a backbone of size $L$ versus $M_3$ for from top
to bottom) $L=16$, 32, 64, 128, 256, and 512. In $M_3$ we count not
virtual bonds but all bonds in the 3-block. The distributions exhibit a
power-law regime with slope $-1.8\pm 0.1$ (b) Distributions scaled with
the value 1.6 for the fractal dimension $d_3$ which gives the best
collapse of the plots in (a).}
\label{pe}
\end{figure}

The fractal dimension of the 3-blocks is considerably smaller than the
fractal dimension, $d_B=1.6432\pm 0.0008$ \cite{Grassberger99}, of
2-blocks (blobs). This is because virtual bonds are counted as one bond
even though they replace many bonds. This can be seen if we plot the
distributions $P^\ast(M_3|L_B)$, the probability that a 3-block
contained in the backbone of a system of size $L$ contains $M_3$ bonds
where we can count not the virtual bonds, but all bonds contained in a
3-block. In Fig.~\ref{pe}(a) we plot $P^\ast(M_3|L_B)$ for various
$L$. The best collapse for these plots (Fig.~\ref{pe}(b)) corresponds to
a fractal dimension of $1.6\pm 0.1$ consistent with the fractal
dimension of 2-blocks in 2D. This can be understood as a reflection of
the fact that in a system of size $L$, the mass of the largest
3-block(counting all bonds) can be the same as the backbone mass. This
is similar to the situation with blobs and backbones; the largest blob
in a backbone can be as large as the whole backbone, which explains why
the fractal dimension of blobs is the same as the fractal dimension of
the backbone.

Replacing a group of bonds by a virtual bond is analogous to
removing dangling ends on a cluster when determining the backbone.

\section{Three Spatial Dimensions}

Our analysis of the results of the 3D simulations proceeds in a similar
manner to the analysis for 2D.

\subsection{3-blocks in Blobs}

\begin{figure}
\centerline{
\xsize
\epsfclipon
\epsfbox{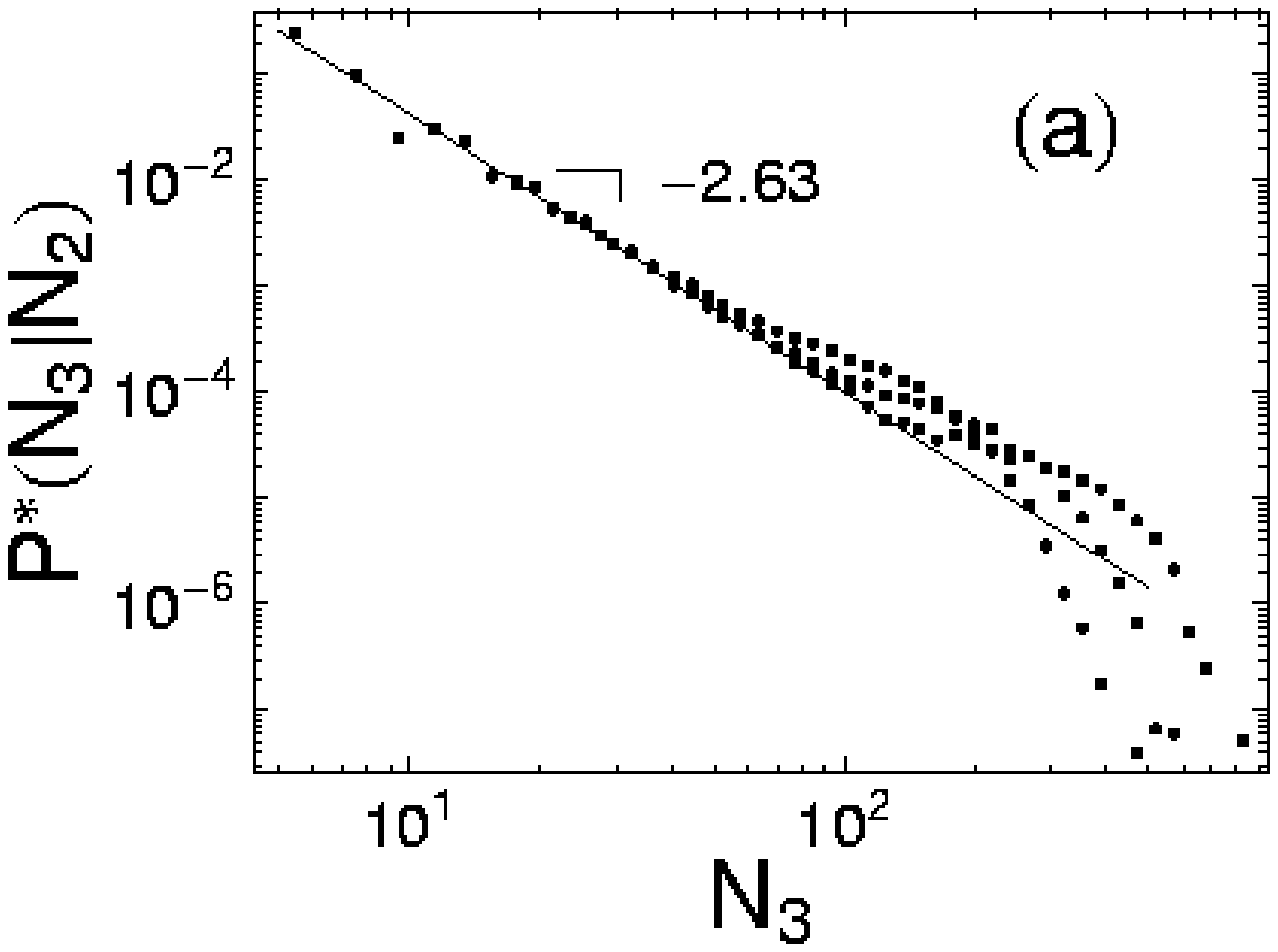}
}
\centerline{
\xsize
\epsfclipon
\epsfbox{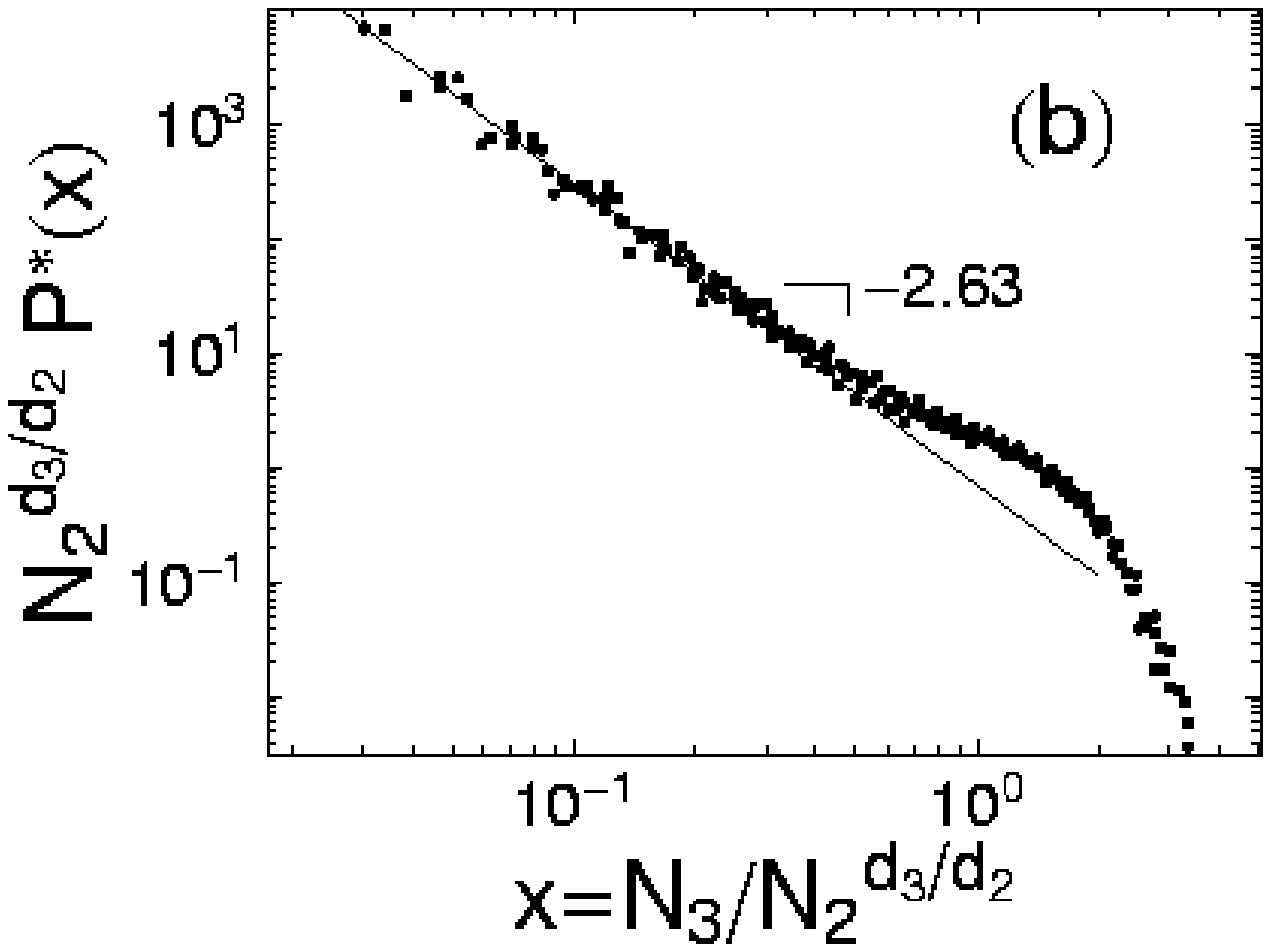}
}
\caption{3D (a) Distributions $P^\ast(N_3|N_2)$ of the number of 3-blocks of
mass $N_3$ in a blob of size $N_2$ versus $N_3$ for (from bottom to top)
$N_2=2^{11}, 2^{12}$, and $2^{13}$. The distributions exhibit a
power-law regime with slope $-2.63\pm 0.05$ (b) Distributions for $N_2=
2^{11}, 2^{12}$, and $2^{13}$ scaled with the value 1.15 for the fractal
dimension $d_3$ which gives the best collapse of the plots in (a).}
\label{pAll33d}
\end{figure}

Figure~\ref{pAll33d}(a) plots the distributions $P^\ast(N_3|N_2)$, the
probability that a 3-block contained in a blob of size $N_2$ contains
$N_3$ bonds, for various values of $N_2$. We estimate the slope of the
power law regimes, $\tau_{3,2}^\ast$, to be $2.63\pm 0.05$.  In
Fig.~\ref{pAll33d}(b), we show the collapsed plots in which we scale the
distributions by ${N_2}^{d_3/d_2}$ with $d_2=1.87\pm 0.03$
\cite{Porto97}.  Visually, we find the best collapse is
obtained for $d_3=1.15\pm 0.1$.

Estimating $d_3$ using Eq.~(\ref{e220}) from the appendix
\begin{equation}
d_{\mbox{\scriptsize 3}}(\tau_{\mbox{\scriptsize
3,2}}^\ast-1)=d_{n2}^\ast=d_2.
\label{e304}
\end{equation}
with $\tau_{\mbox{\scriptsize 3,2}}^\ast=2.63\pm 0.05$ and $d_2=1.87\pm
0.03$, results in an estimate of $d_3=1.15\pm 0.05$.

\subsection{3-blocks in Backbone}

\begin{figure}
\centerline{
\xsize
\epsfclipon
\epsfbox{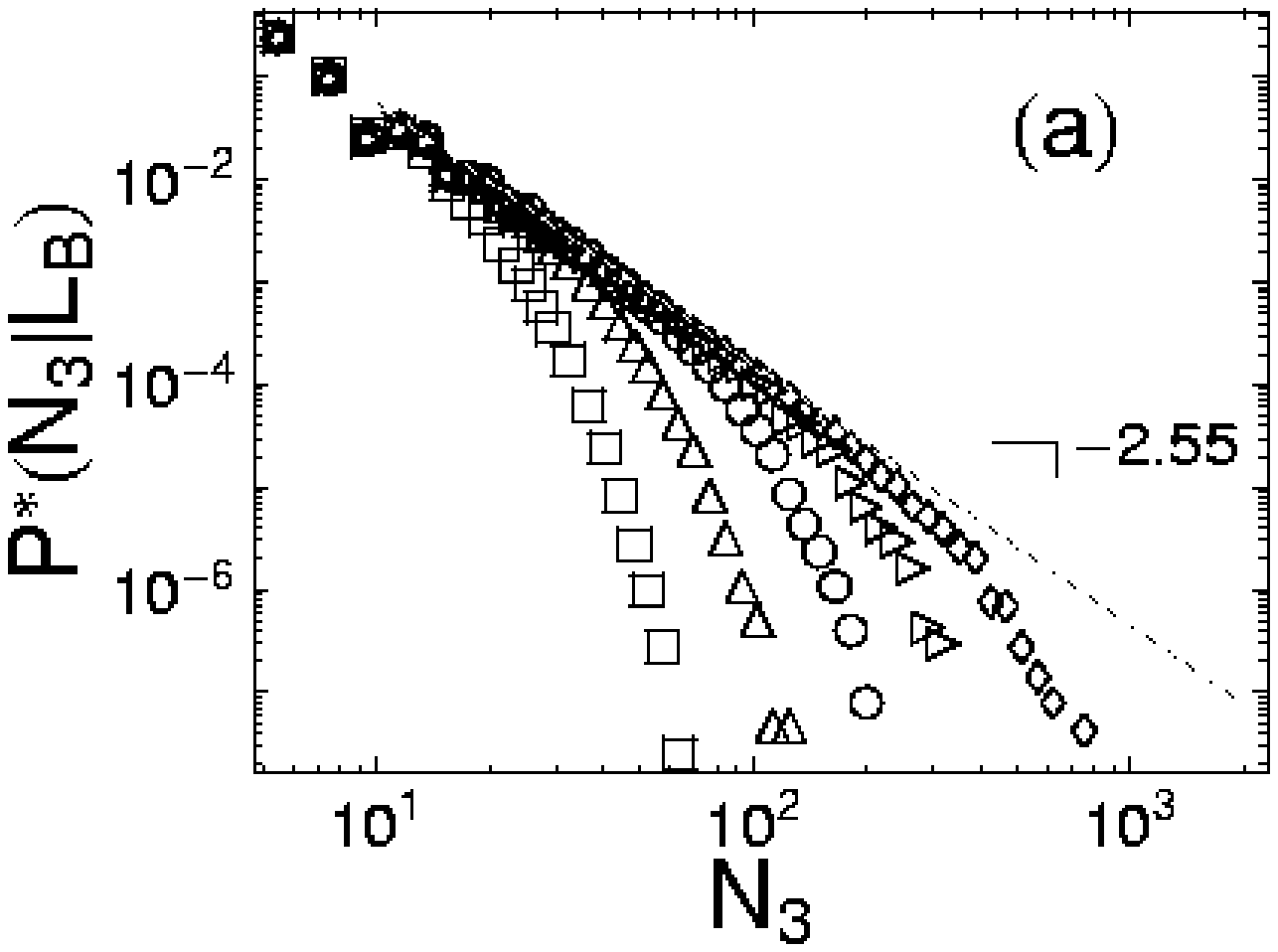}
}

\centerline{
\xsize
\epsfclipon
\epsfbox{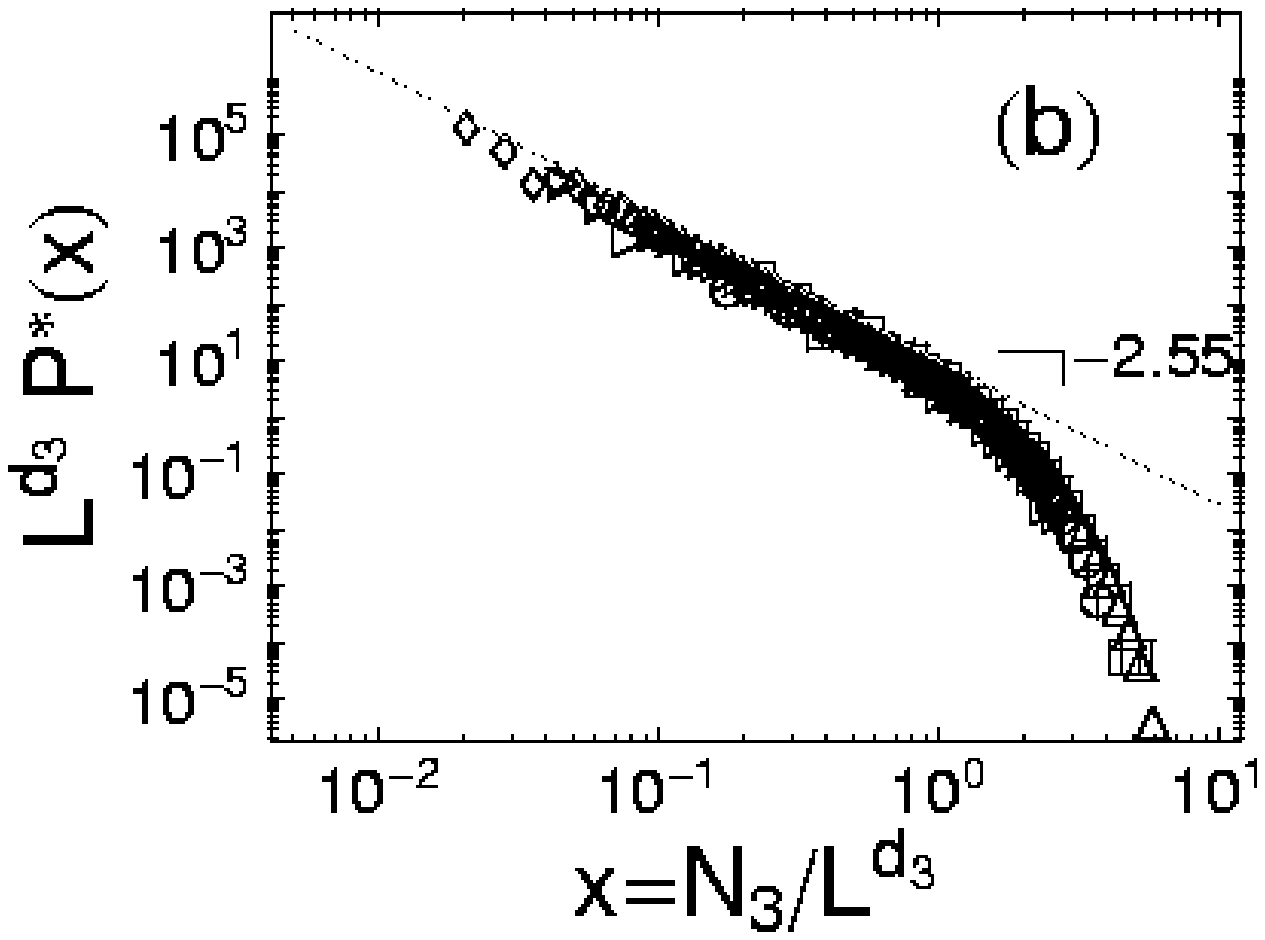}
}

\caption{3D (a) Distributions $P^\ast(N_3|L)$ of the number of 3-blocks of
mass $N_3$ in a backbone of size $L$ versus $N_3$ for (from top to bottom)
$L= 8$, 16, 32, 64, and 128. The distributions exhibit a power-law
regime with slope $-2.55\pm 0.1$. (b) Distributions scaled with the
value 1.15 for the fractal dimension $d_3$ which gives the best collapse
of the plots in (a).}
\label{pc3d}
\end{figure}

Figure~\ref{pc3d}(a) plots the distributions $P^\ast(N_3|L_B)$, the
probability that a 3-block contained in the backbone of a system of size
$L$ contains $N_3$ bonds, for various values of $L$.  We estimate
the slope of the power law regimes, $\tau_{3,B}^\ast$, to be $2.55\pm
0.05$. In Fig.~\ref{pc3d}(b), we show the collapsed plots in which we scale
the distributions by $L^{d_3}$. Visually, we find the best collapse is
obtained for $d_3=1.15\pm 0.1$.

\begin{figure}
\centerline{
\xsize
\epsfclipon
\epsfbox{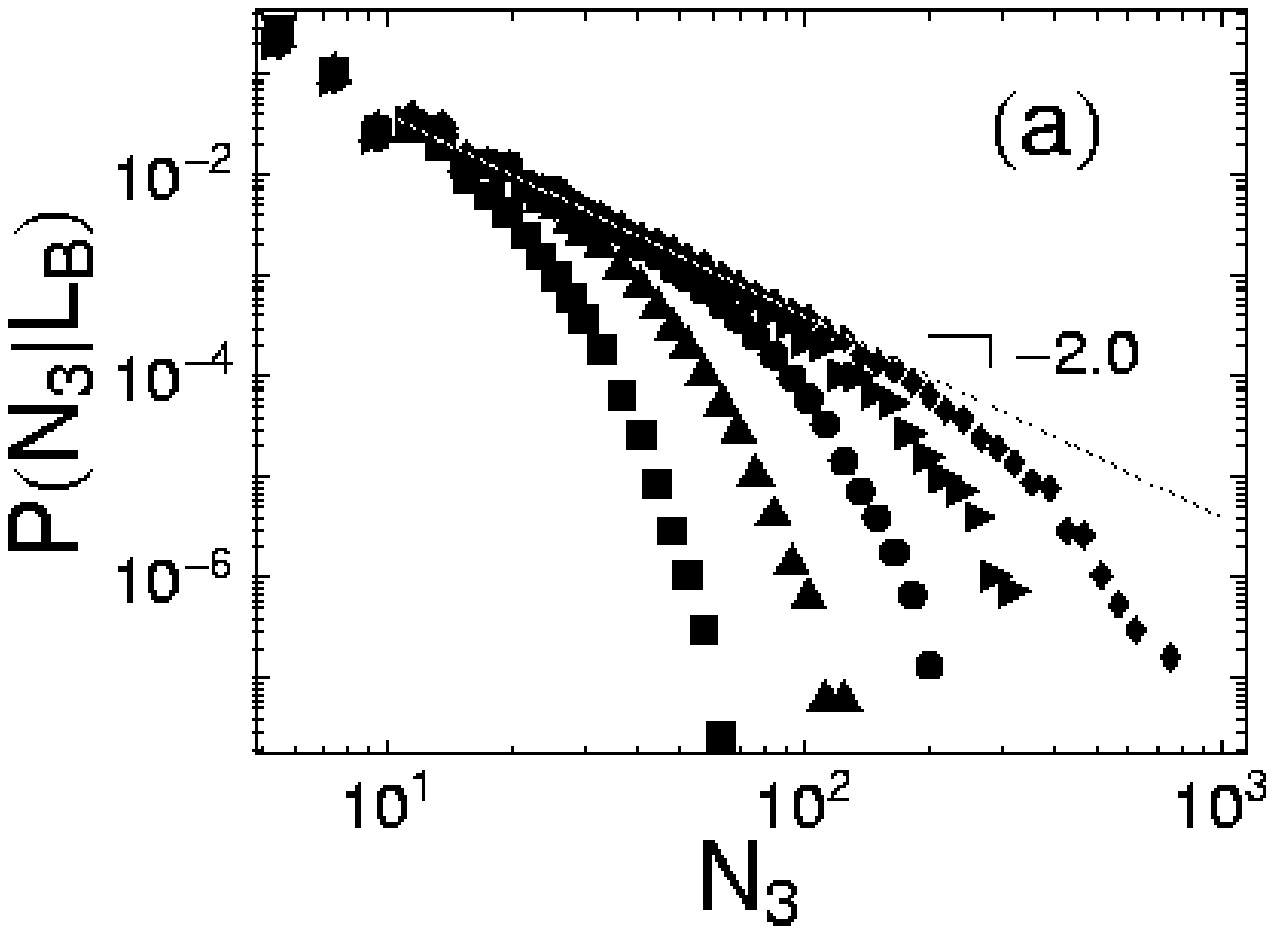}
}
\centerline{
\xsize
\epsfclipon
\epsfbox{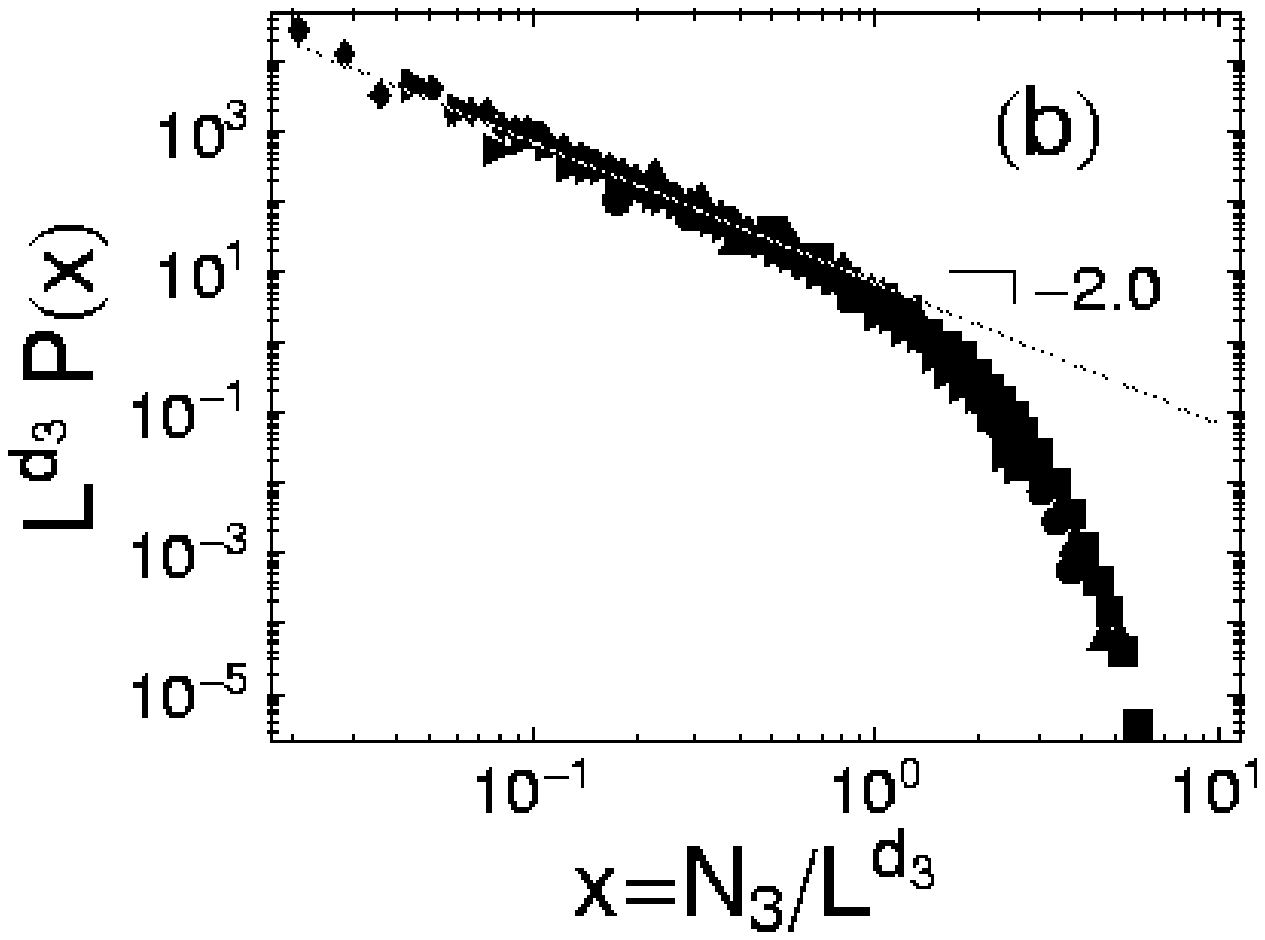}
}
\caption{3D (a) Distributions $P(N_3|L)$ of the number of top level
3-blocks of mass $N_3$ in a backbone of size $L$ versus $N_3$ for (from
top to bottom) $L=32$, 64, 128, 256, and 512. The distributions exhibit a
power-law regime with slope $-2.0\pm 0.1$. (b) Distributions scaled with
the value 1.15 for the fractal dimension $d_3$ which gives the best
collapse of the plots in (a).}
\label{pc13d}
\end{figure}

\begin{figure}

\centerline{
\xsize
\epsfclipon
\epsfbox{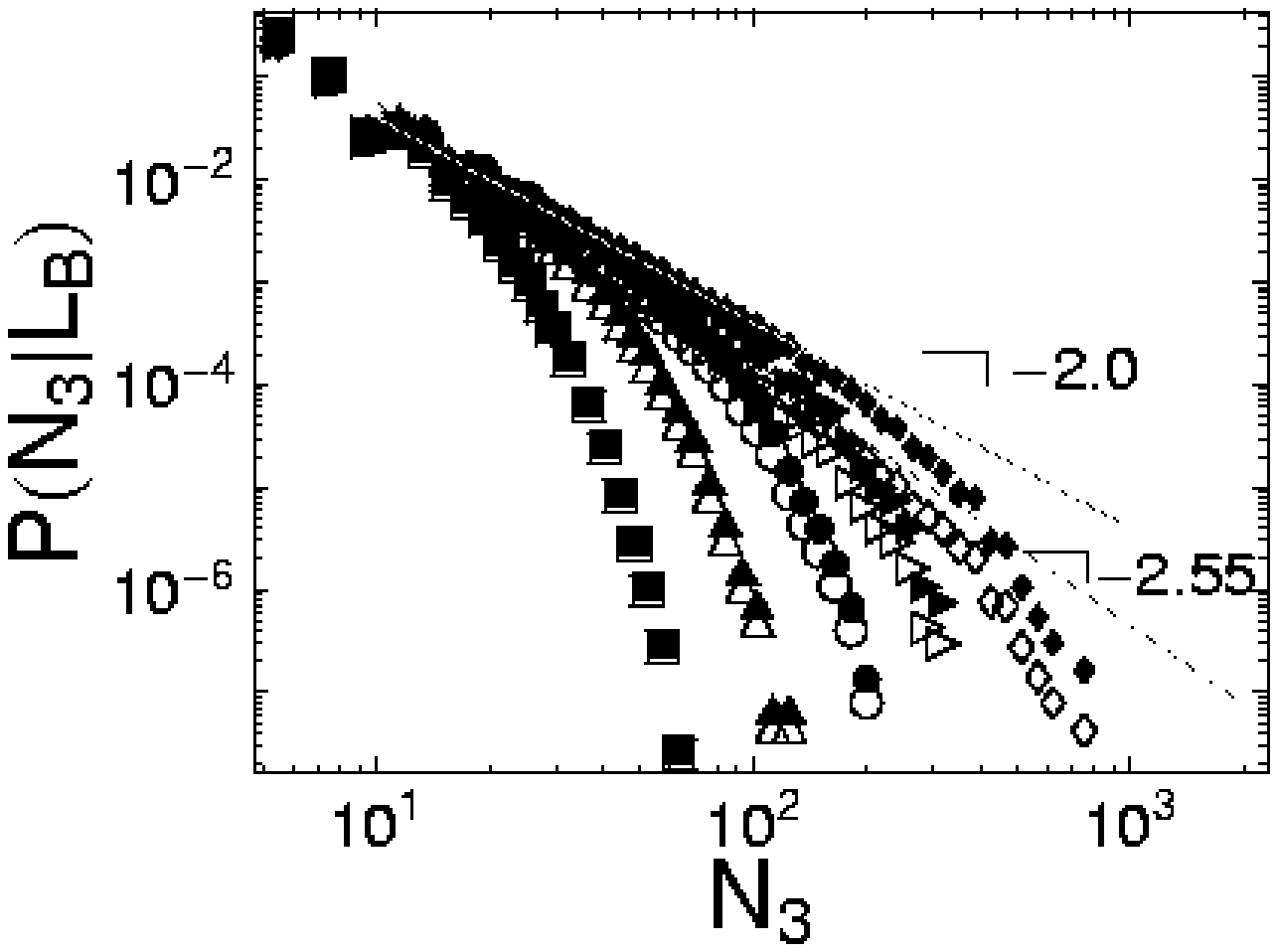}
}
\caption{3D Distributions $P(N_3|L)$ of top level 3-blocks (filled
symbols) and $P^\ast(N_3|L)$ of all-level 3-blocks (unfilled symbols). While
the slopes of the power law regimes of the two types of distributions
are different, the finite-size-system cutoffs are essentially
superimposed consistent, with the fractal dimension of the two types of
distributions being equal.}
\label{pcComb13d}
\end{figure}

Next we consider the distribution of ``top-level'' 3-blocks in the
backbone.  In Fig.~\ref{pc13d}(a), we plot the distributions $P(N_3|L_B)$,
the probability that a top-level 3-block contained in the backbone of a
system of size $L$ contains $N_3$ bonds, for various values of $L$.  The
exponent of the power-law regimes $\tau_{3,B}$ is estimated to be
$2.0\pm 0.05$.  In Fig.~\ref{pc13d}(b), we show the collapsed plots, in
which we scale the distributions by $L^{d_3}$. The best collapse is
obtained for $d_3=1.15\pm 0.1$, the same value as for the distributions
of 3-blocks of all levels. As in 2D, the fractal dimensions of the top level
3-blocks is the same as the fractal dimension of 3-blocks of all levels
but the slopes of the power law regimes are different; this is seen also
in Fig.~\ref{pcComb13d}.

Using Eq.~(\ref{e210})
\begin{equation}
d_{\mbox{\scriptsize 3}}(\tau_{\mbox{\scriptsize
3,B}}-1)=d_{nB}={1\over\nu}.
\label{e306}
\end{equation}
to obtain an estimate of $d_3$ with our estimate for $\tau_{3,B}$ above
we find $d_3=1.14\pm 0.1$. Combining this result with our earlier
estimates, we make the final estimate
\begin{equation}
d_3=1.15\pm 0.1.
\label{e9}
\end{equation}

The simulation results notwithstanding, it would be surprising if $d_3$
were smaller in 3D than in 2D because, below the critical dimension
$d_c=6$, both the fractal dimensions of clusters and blobs increase with
the Euclidean dimension. This suggests that while the actual values of
$d_3$ may be within the bounds we have estimated, the actual values will
be consistent with $d_3~(\mbox{2D})\leq d_3~(\mbox{3D})$.

\begin{figure}
\centerline{
\xsize
\epsfclipon
\epsfbox{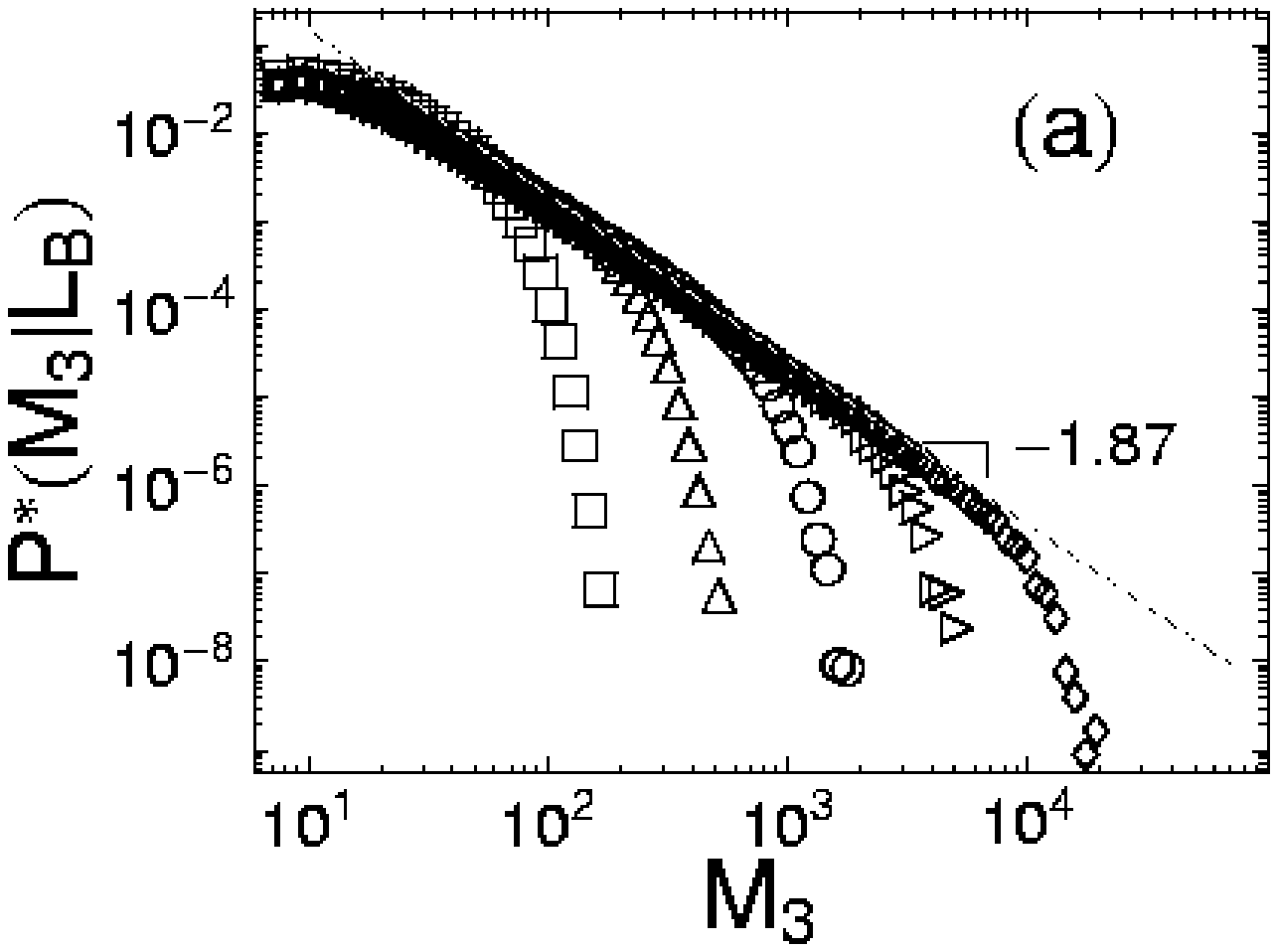}
}
\centerline{
\xsize
\epsfclipon
\epsfbox{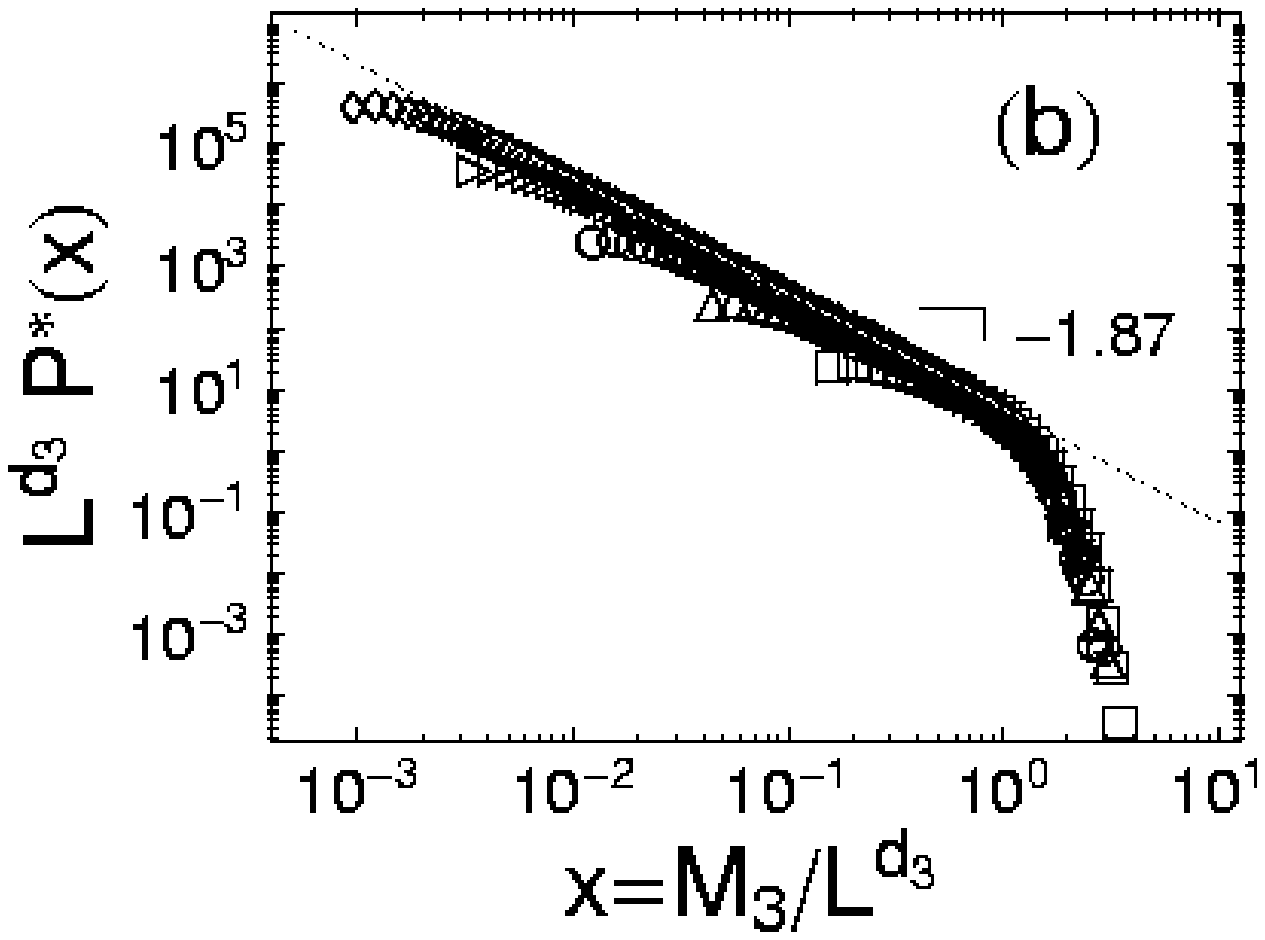}
}
\caption{3D (a) Distributions $P^\ast(M_3|L)$ of the number of 3-blocks of
mass $M_3$ in a backbone of size $L$ versus $M_3$ for (from top to bottom)
$L=8$, 16, 32, 64, and 128. In $M_3$ we count not virtual bonds but all
bonds in the 3-block. The distributions exhibit a power-law regime with
slope $-1.87\pm 0.1$. (b) Distributions scaled with the value 1.85 for
the fractal dimension $d_3$ which gives the best collapse of the plots
in (a).}
\label{pe3d}
\end{figure}

As in 2D, if we do not replace two-terminal objects in a 3-block by a
single virtual bond, the fractal dimension of the 3-block is that of a
blob(see Fig. \ref{pe3d}).

Estimates for all of the 2D and 3D exponents are summarized in Table I.

\section{Decomposition of the Whole Percolation Cluster}

While we have only decomposed 2-blocks that comprise the cluster
backbone, we could proceed similarly for all 2-blocks into which a
cluster is decomposed. The fractal dimension of the 3-blocks into which
a cluster is ultimately decomposed should be the same as the fractal
dimension of the 3-blocks into which the backbone is ultimately
decomposed. The only difference we would expect in our results would
be that the slope of the power-law regime of the distribution of
top-level 3-blocks would be given by
\begin{equation}
d_{\mbox{\scriptsize 3}}(\tau_{\mbox{\scriptsize{
3,1}}}-1)=d_{n1}=d.
\label{e209}
\end{equation}
the analogy of Eq.~(\ref{e210})

\section{$k$-Bones and Path Crossing Probabilities} 

Just as blobs and backbones have the same fractal dimension, we can
identify objects analagous to backbones which have the same fractal
dimensions as $k$-blocks. We define a $k$-bone as the set of all points
in a percolation system connected to $k$ disjoint terminal points(or
sets of disjoint terminal points) by $k$ disjoint paths. Thus the
backbone is a $k$-bone with $k=2$. Just as the largest $k$-blocks into
which a backbone can be decomposed are 2-blocks, the largest $k$-blocks
into which a $k$-bone can be decomposed are $k$-blocks. The fractal
dimension of $k$-bones is the fractal dimension of the $k$-blocks. One
can see this easily by noting that if the $k$ terminal points which
define a $k$-bone are connected to each other, the resulting structure
is $k$-block.

Recent work \cite{AizenmanXX} has identified a relationship between path
crossing probabilities and the fractal dimensions of percolation
structures. Specifically, consider the probability, $\hat P_k^P$ that in
an annular region the small inner circle of radius $r$ is connected to
the larger outer circle of radius $R$, $R\gg r$, by $k$ disjoint
paths. Then 
\begin{equation}
\hat P_k^P\sim\left({r\over R}\right)^{\hat x_k}.
\label{e14a}
\end{equation}
It has been observed \cite{AizenmanXX} that $\hat x_1$ is the codimension
of the percolation cluster and $\hat x_2$ is the codimesion of the
backbone. We extend these observations to the case of general $k$
\begin{equation}
d-\hat x_k=d_k,
\label{e14b}
\end{equation}
where $d$ is the spatial dimension of the system. This should hold in
all dimensions where the annulus is now defined by two hyperspheres. It
has been argued \cite{JacobsenXX} that
\begin{equation}
x_k<\hat x_k<x_{2k},
\label{e14c}
\end{equation}
where $x_k$ is the polychromatic path crossing exponent
\cite{AizenmanXX} and which has been found rigorously in 2D to be
\cite{AizenmanXX}
\begin{equation}
x_k={1\over 12}(k^2-1).
\label{e14d}
\end{equation}
Using Eqs.~(\ref{e14b}), (\ref{e14c}), and (\ref{e14d}), we find in 2D
\begin{equation}
-{11\over 12}<d_3<{4\over 3},
\label{e14e}
\end{equation}
consistent with our estimate for $d_3$ in 2D. 

The relationship between the path crossing problem for $k=2$ and the
backbone dimension has been recently exploited to determine $d_B$ very
accurately using a transfer matrix technique \cite{JacobsenXX}. Possibly
similar methods can be employed to find the fractal dimension of
$k$-bones (and therefore $k$-blocks) with $k\geq 3$ to high precision.

\section{Relationship to Renormalization Group}

The process of replacing a 2-terminal object, $t$, by a single virtual
bond and then replacing 2-terminal objects within $t$ by single virtual
bonds and so on is reminiscent of the decimation process in
renormalization group (RG) approaches to percolation
\cite{Stauffer94,Bunde96,Hovi,Reynolds}. It is here, however, that the
similarity ends. The decimation process performed in the decomposition
into 3-blocks is an exact decimation performed on objects in individual
realizations while the RG decimation is performed on the lattice and is
an approximation, except for hierarchical lattices. Also, the purpose of
the decomposition into 3-blocks is to improve computational performance
and analyze the properties of substructures of the cluster while the
purpose of RG calculations is to find properties of percolation
analytically. Finally, whereas RG approaches on hierarchical lattices
result in objects that are finitely ramified, the decomposition into
3-blocks we perform maintains the infinite ramification of the Euclidean
lattice.

\section{Computational Implications}

The fact that the fractal dimension of 3-blocks is significantly smaller
than the fractal dimension of 2-blocks has important computational
implications. We can efficiently calculate properties(e.g. resistance,
velocity distributions, self avoiding walk statistics) of a percolation
cluster or backbone as follows:

\begin{itemize}

\item[{(i)}] decompose the cluster or backbone into 2-blocks
\item[{(ii)}] decompose the 2-blocks into 3-blocks
\item[{(iii)}] calculate the desired properties of the 3-blocks
\item[{(iv)}] algebraically determine the properties of the 2-blocks from the
properties of the 3-blocks
\item[{(v)}] algebraically determine the properties of the cluster or backbone
from the properties of the 2-blocks.

\end{itemize}

In many cases the computation will require less CPU (computer
processing) resource when the complexity of the computation is a power
law or exponential of the mass of the object for which the property is
being calculated. By decomposition we make the mass of these objects
smaller. Reduced CPU resource usage is also obtained if only a
decomposition into 2-blocks is made although the saving are
less. Systems of larger size than could be treated before can now be
treated when we decompose into 3-blocks because the fractal dimension of
the 3-blocks are lower than that of the object in which they are
embedded; this is not true if we only decompose into 2-blocks.

As an example of the dramatically smaller size of the largest 3-block
versus the size of the largest blob consider a 3D system of size
$L=1000$. At criticality, the largest mass blob in the backbone will be
of the order $L^{1.62}\approx 63,000$ while the mass of the largest
3-block in the backbone will be of $L^{1.2}\approx 4000$.  In
Fig. \ref{preal} we show an actual simulation realization in which a
blob of 950 bonds is decomposed into a 3-block with only 216 virtual
bonds, greatly reducing the computational complexity.

\section{Discussion}

Traditionally the decomposition of percolation systems has been to
decompose the system into clusters (1-blocks) and to decompose the
clusters into blobs (2-blocks). We extend this decomposition by
decomposing 2-blocks into 3-blocks. 3-blocks are especially interesting
because in contrast to 1- and 2-blocks, the 3-blocks have the property
that they can be nested. That is, 2-terminal objects, which are replaced
by single virtual bonds in a 3-block, can themselves contain other
3-blocks. Because of this replacement of a 2-terminal object by a
virtual bond, the fractal dimension of 3-blocks is significantly smaller
than the fractal dimension of 2-blocks. As discussed in the previous
section, this smaller fractal dimension has important computational
implications for the size of percolation systems which can be analyzed
and the speed at which the analysis can be performed.

In addition, within the error bars of our calculations, the values for
the 3-block fractal dimension appear to be identical for 2D and 3D
systems. Simulations of larger systems and higher dimension systems
could help answer whether in fact $d_3$ is independent of dimension
(``super-universal''). It will also be of interest to determine the
properties of $k$-blocks with $k>3$.

\subsubsection*{Acknowledgements}

We thank D. Baker, L. Braunstein, S. Havlin, A. Moreira, and
V. Schulte-Frohlinde for stimulating discussions.

\appendix
\section{Relationships among Exponents}

Here we ask if any of the fractal dimensions and power-law regime
exponents we have identified are related. To answer this question we
must first briefly review some existing results for relations between
other exponents.

\subsection{Previous Results}

It has been shown generally \cite{Huber95a,Huber95b} that, for disjoint
objects of type $X$ embedded in a space $Y$,
\begin{equation}
d_X(\tau_{X,Y}-1)=d_{nY}.
\label{e3}
\end{equation}
Equation (\ref{e3}) holds if $\tau_X<2$ or if $d_{nY}$ is equal to the
fractal dimension of space $Y$, $d_Y$.

Special cases of Eq.~(\ref{e3}) have been identified previously for
$Y=0$, 1, and 2 corresponding to Euclidean space, percolation cluster
space and percolation backbone space, respectively.

\begin{itemize}
\item[{(i)}] The first is the familiar scaling relation for the Fisher exponent
$\tau$ \cite{Stauffer94,Bunde96}
\begin{equation}
d_f(\tau-1)=d_{n0}=d \qquad\qquad\mbox{clusters},
\label{e4a}
\end{equation}
where $d$ is the Euclidean dimension, $d_f$ the fractal dimension of the
cluster, and $\tau$ the exponent of the power-law regime in the
distribution of cluster sizes.

\item[{(ii)}] In Ref.~\cite{Gyure95} it was shown that
\begin{equation}
d_{\mbox{\scriptsize blob-cl}}(\tau_{\mbox{\scriptsize
blob-cl}}-1)=d_{n1}=d_f, \qquad \mbox{(cluster blobs)}
\label{e5}
\end{equation}
where $d_{\mbox{\scriptsize blob-cl}}$ and $\tau_{\mbox{\scriptsize
blob-cl}}$ are the fractal dimension and the exponent of the power-law
regime, respectively, for all blobs in the cluster.

\item[{(iii)}] In Ref.~\cite{Herrmann84} it was argued that
\begin{equation}
d_{\mbox{\scriptsize blob-bb}}(\tau_{\mbox{\scriptsize
blob-bb}}-1)=d_{nB}=d_{\mbox{\scriptsize red}}, \quad
\mbox{(backbone blobs)}
\label{e6}
\end{equation}
where $d_{\mbox{\scriptsize blob-bb}}$ and $\tau_{\mbox{\scriptsize
blob-bb}}$ are the fractal dimension and the exponent of the power-law
regime, respectively, for those blobs in the backbone and
$d_{\mbox{\scriptsize red}}$ is the fractal dimension of
singly-connected red bonds in the backbone. 

\end{itemize}

Both $d_{\mbox{\scriptsize blob-cl}}$ and $d_{\mbox{\scriptsize
blob-bb}}$ are equal to $d_B$, the backbone fractal dimension.
In (i) and (ii), Eq.~(\ref{e3}) applies because $d_{nY}=d_Y$; in (iii),
Eq.~(\ref{e3}) applies because $\tau_X<2$.

\subsection{3-blocks in Blobs}

\begin{figure}
\centerline{
\xsize
\epsfclipon
\epsfbox{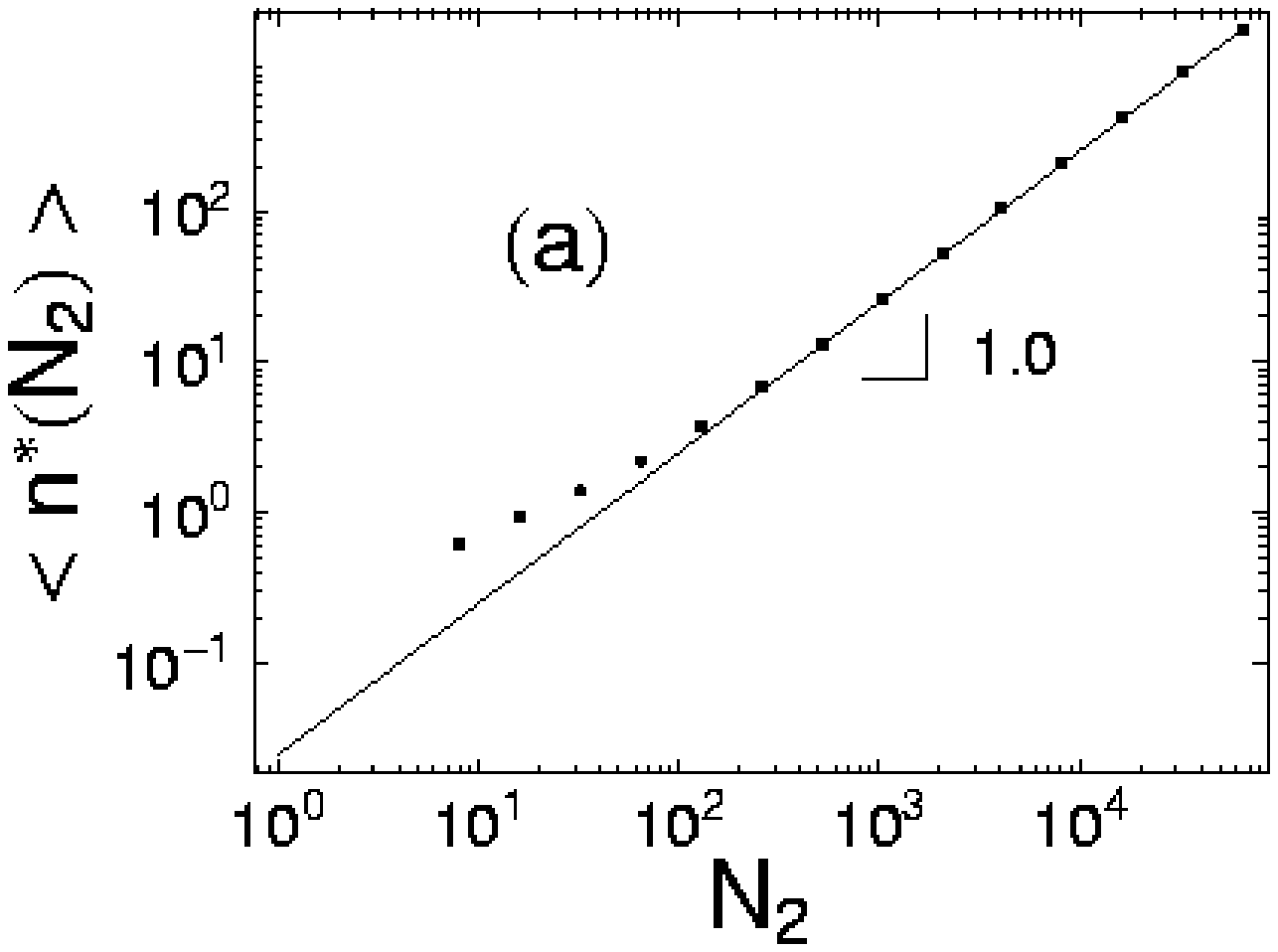}
}

\centerline{
\xsize
\epsfclipon
\epsfbox{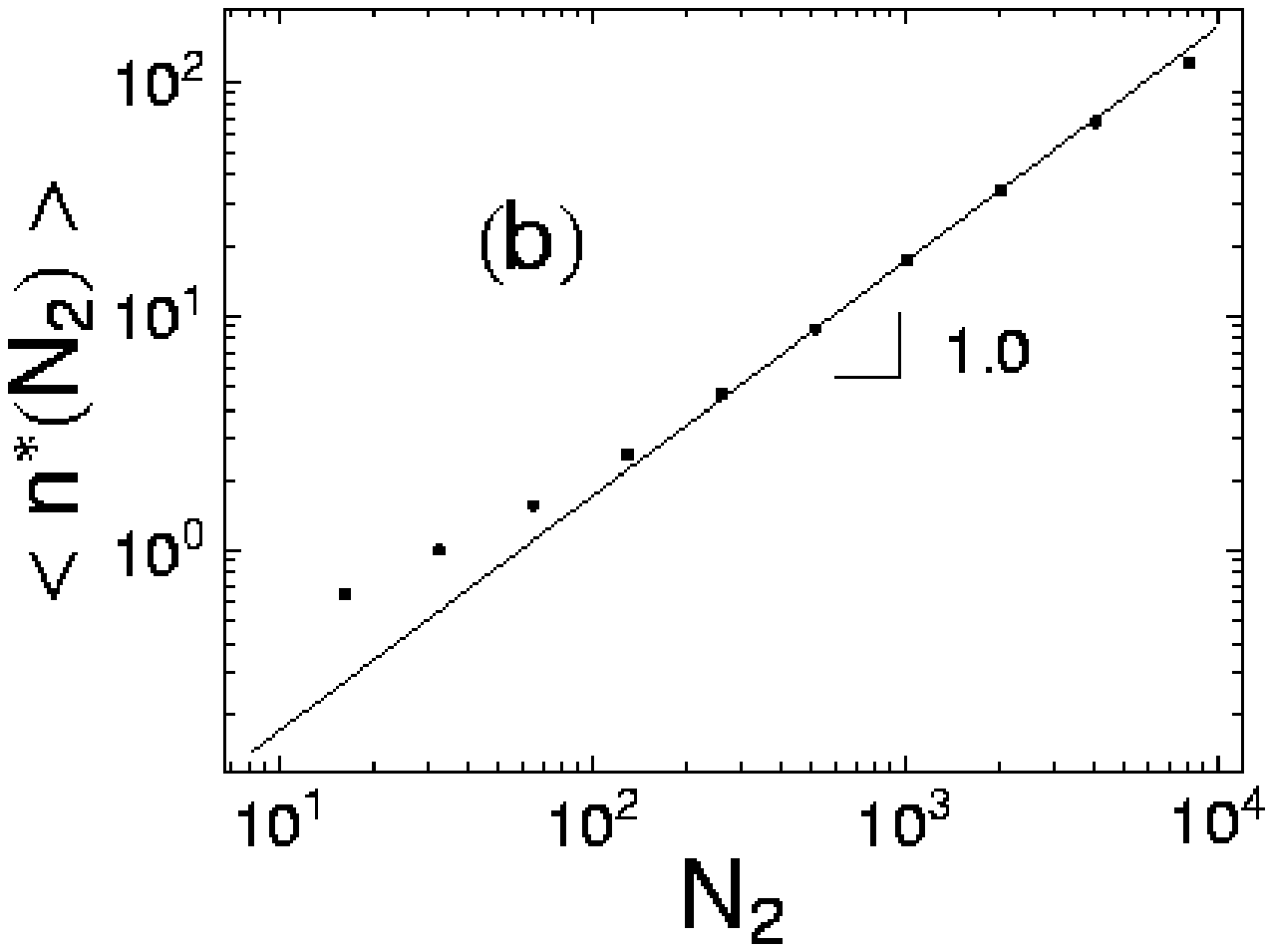}
}
\caption{$\langle n^\ast(N_2)\rangle$, the average number of 3-blocks in
a blob of size $N_2$ versus $N_2$ for (a) 2D and (b) 3D.}
   
\label{pNumbAll3}
\end{figure}

In analogy with Eqs.~(\ref{e4a}) and~(\ref{e5}), we would expect
\begin{equation}
d_{\mbox{\scriptsize 3}}(\tau_{\mbox{\scriptsize
3,2}}^\ast-1)=d_{n2}^\ast=d_2.
\label{e220}
\end{equation}
We first confirm that the total number of 3-blocks in blobs scales with
the exponent $d_2$.  If
\begin{equation}
<n(L)>\sim L^{d_{n2}}
\label{e206}
\end{equation}
and
\begin{equation}
N_2\sim L^{d_2}
\label{e207}
\end{equation}
then we would expect
\begin{equation}
<n(N_2)>\sim L^{d_{n2}/d_2}.
\label{e208}
\end{equation}
Figures \ref{pNumbAll3}(a) and \ref{pNumbAll3}(b) are log-log plots of
$\langle n(N_2)\rangle$, the average number of all 3-blocks in a
blob, versus blob size $N_2$ for 2D and 3D, respectively. The straight
line fits with slope $1.0\pm 0.05$ are consistent with $d_{n2}=d_2$.
Our simulation results in 2D from Section IV, $d_3=1.20$ and
$\tau_{3,2}=2.35$ result in $d_3(\tau_{3,2}^\ast-1)=1.62$ close to the value
$d_2=1.6432$.  In 3D, our simulation results from Section V, $d_3=1.15$ and
$\tau_{3,2}=2.63$ result in $d_3(\tau_{3,2}^\ast-1)=1.87$ identical to the value
$d_2=1.87$\cite{Porto97}.

\subsection{3-blocks in Backbone}

Because the number of top-level 3-blocks in the backbone is proportional
to the number of 2-blocks in the backbone, the number of top level 3-
blocks in the backbone should scale the same way the number of 2-blocks
in the backbone.  For all dimensions and lattices, $d_{nB}$ has been
shown to be \cite{Coniglio82a,Coniglio82b}
\begin{equation}
d_{nB}=d_{red}={1\over\nu},
\label{e7}
\end{equation}
where $\nu$ is the exponent associated with the divergence of the
correlation length as $p$ approaches $p_c$ [1,2]. In 2D $1/\nu$ is
exactly 3/4 \cite{denNijs82,Nienhuis82}; in 3D, $1/\nu$ has been
estimated to be $1.143\pm 0.01$ \cite{ZiffStell,Strenski}.  We would
expect
\begin{equation}
d_{\mbox{\scriptsize 3}}(\tau_{\mbox{\scriptsize
3,B}}-1)=d_{nB}={1\over\nu}.
\label{e210}
\end{equation}

\begin{figure}
\centerline{
\xsize
\epsfclipon
\epsfbox{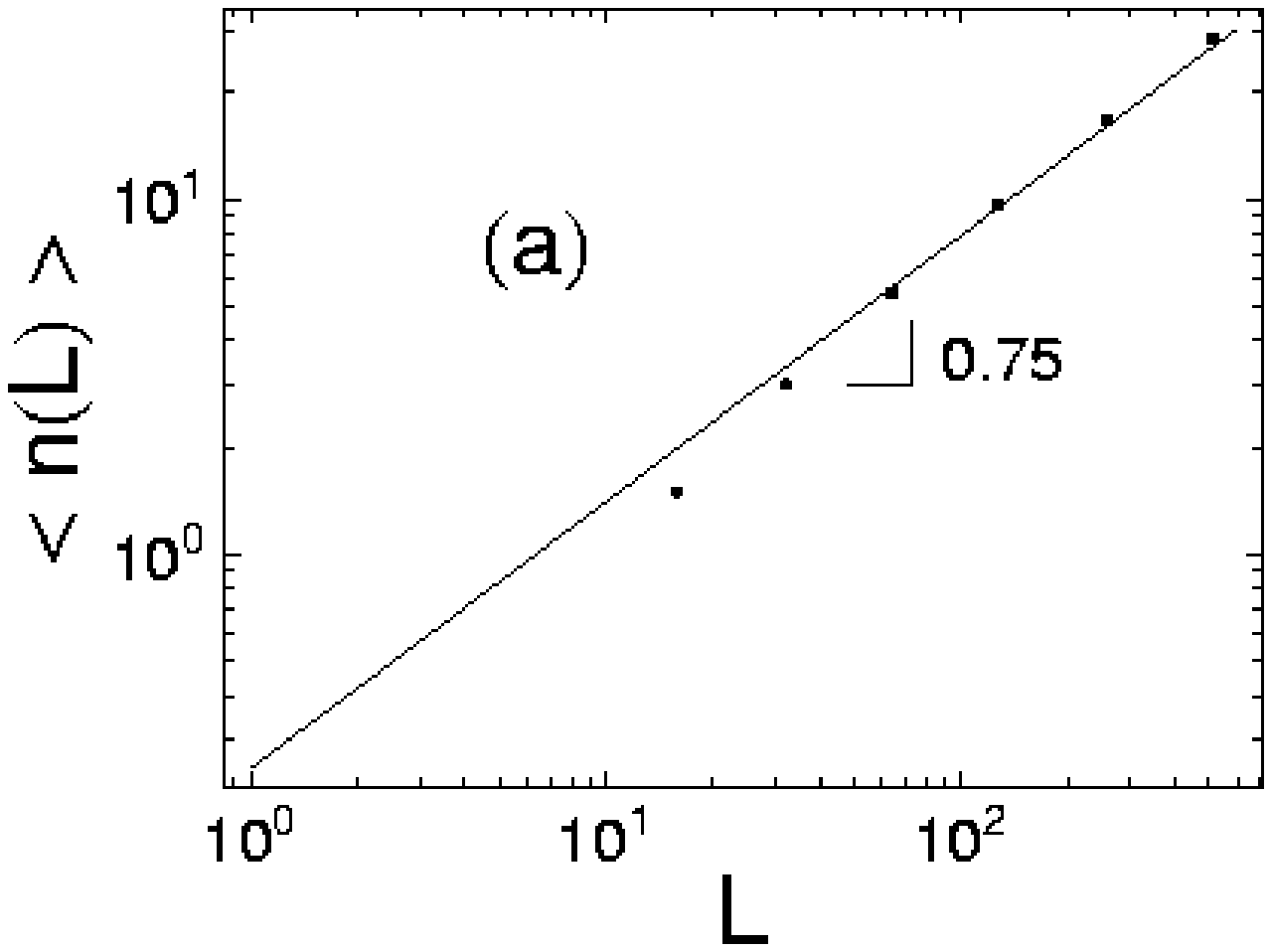}
}

\centerline{
\xsize
\epsfclipon
\epsfbox{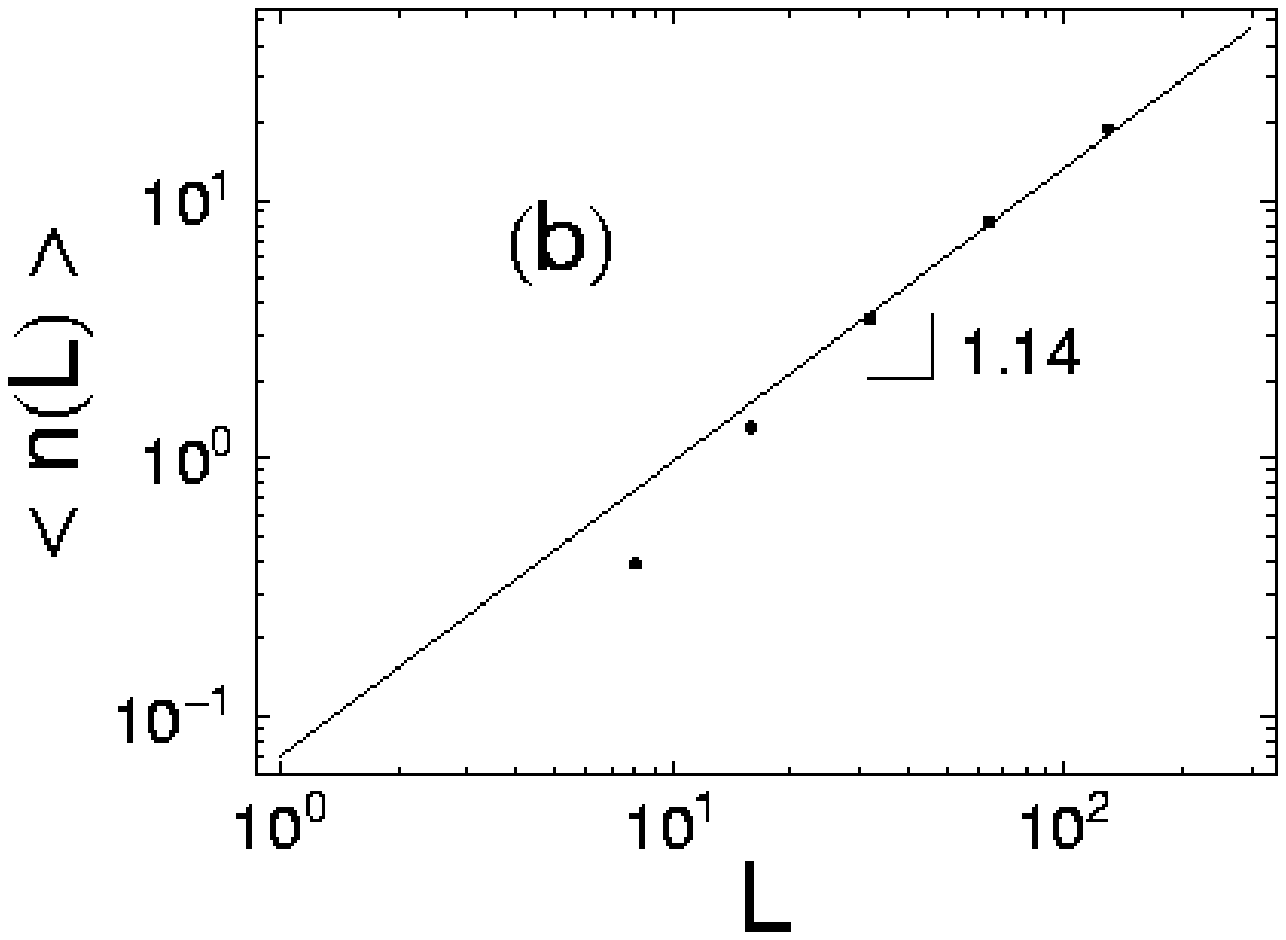}
}

\caption{$\langle n(L)\rangle$, the average number of top level
3-blocks in a backbone of size $L$ versus $L$. (a) 2D The solid line has
slope 0.75. (b) 3D The solid line has slope 1.14.}
\label{pNumbC}
\end{figure}

Figures \ref{pNumbC}(a) and \ref{pNumbC}(b) are log-log plots of $\langle
n(L)\rangle$, the average number of top-level 3-blocks in the backbone,
versus system size $L$ for 2D and 3D, respectively. The straight line
fits with slope $0.75\pm 0.05$ and $1.14\pm 0.05$ are consistent with
the exact and previously estimated values for $1/\nu$ of 3/4 and 1.143
in 2D and 3D, respectively.  Our 2D simulation results from Section IV,
$d_3=1.15$ and $\tau_{3,2}=1.60$ result in $d_3(\tau_{3,B}-1)=0.69$
close to the value $1/\nu=3/4$.  For 3D, our simulation results from
Section V, $d_3=1.15$ and $\tau_{3,2}=2.0$ result in
$d_3(\tau_{3,B}-1)=1.15$ close to the value $1/\nu=1.143$.

\end{multicols}

\newpage

%%%%%%%%%%%%%%%%%%%%%%%%%%%%%%%%%%%%%%%%%%%%%%%%%%%%%%%%%%%%%%%%%%%%%

\begin{table}[htb]
\caption{Measured fractal dimension, measured power-law regime exponent,
and calculated fractal dimension  for 3-blocks in 2D and 3D. The
calculated value of $d_3$ is determined by Eq.~(\protect\ref{e104}) for 3-blocks in a blob
and Eq.~(\protect\ref{e106}) for 3-blocks in the backbone.}
\begin{tabular}{cccc}
{\bf 2D}\\
           & $d_3$ & $\tau$ & $d_3$ \\ 
           & {\scriptsize MEASURED} & {\scriptsize MEASURED} 
& {\scriptsize CALCULATED} \\ \hline
{\scriptsize All 3 blocks in blob} & $1.20\pm 0.1$ & $2.35\pm 0.05$ &
             $1.22\pm 0.05$ \\
{\scriptsize All 3 blocks in backbone} & $1.15\pm 0.1$ & $2.25\pm 0.05$ &
          --- \\
{\scriptsize Top level 3 blocks in backbone} & $1.15\pm 0.1$ & $1.60\pm 0.05$ &
           $1.25\pm 0.1$ \\  \hline
{\bf 3D}\\
           & $d_3$ & $\tau$ & $d_3$ \\
           & {\scriptsize MEASURED} & {\scriptsize MEASURED}
 & {\scriptsize CALCULATED} \\ \hline
{\scriptsize All 3 blocks in blob} & $1.15\pm 0.1$ & $2.63\pm 0.05$ &
            $1.15\pm 0.05$ \\
{\scriptsize All 3 blocks in backbone} & $1.15\pm 0.1$ & $2.55\pm 0.05$
           &
           --- \\
{\scriptsize Top level 3 blocks in backbone} & $1.15\pm 0.1$ & $2.0\pm
           0.05$ &
            $1.14\pm 0.1$ \\
\end{tabular}
\end{table}

%%%%%%%%%%%%%%%%%%%%%%%%%%%%%%%%%%%%%%%%%%%%%%%%%%%%%%%%%%%%%%%%%%%%%

%%%%%%%%%%%%%%%%%%%%%%%%%%%%%%%%%%%%%%%%%%%%%%%%%%%%%%%%%%%%%%%%%%%%%

\newpage 

%%%%%%%%%%%%%%%%%%%%%%%%%%%%%%%%%%%%%%%%%%%%%%%%%%%%%%%%%%%%%%%%%%%%
%  real lattice/tpos/3block
%%%%%%%%%%%%%%%%%%%%%%%%%%%%%%%%%%%%%%%%%%%%%%%%%%%%%%%%%%%%%%%%%%%%%
\newpage 

%%%%%%%%%%%%%%%%%%%%%%%%%%%%%%%%%%%%%%%%%%%%%%%%%%%%%%%%%%%%%%%%%%%%%

\newpage 

%%%%%%%%%%%%%%%%%%%%%%%%%%%%%%%%%%%%%%%%%%%%%%%%%%%%%%%%%%%%%%%%%%%%%

\newpage

%%%%%%%%%%%%%%%%%%%%%%%%%%%%%%%%%%%%%%%%%%%%%%%%%%%%%%%%%%%%%%%%%%%%%

\newpage

%%%%%%%%%%%%%%%%%%%%%%%%%%%%%%%%%%%%%%%%%%%%%%%%%%%%%%%%%%%%%%%%%%%%%

\newpage

%%%%%%%%%%%%%%%%%%%%%%%%%%%%%%%%%%%%%%%%%%%%%%%%%%%%%%%%%%%%%%%%%%%%%
%  3D
%%%%%%%%%%%%%%%%%%%%%%%%%%%%%%%%%%%%%%%%%%%%%%%%%%%%%%%%%%%%%%%%%%%%%

\newpage 

%%%%%%%%%%%%%%%%%%%%%%%%%%%%%%%%%%%%%%%%%%%%%%%%%%%%%%%%%%%%%%%%%%%%%
\newpage

%%%%%%%%%%%%%%%%%%%%%%%%%%%%%%%%%%%%%%%%%%%%%%%%%%%%%%%%%%%%%%%%%%%%%

\newpage

%%%%%%%%%%%%%%%%%%%%%%%%%%%%%%%%%%%%%%%%%%%%%%%%%%%%%%%%%%%%%%%%%%%%%

\newpage

\newpage

%%%%%%%%%%%%%%%%%%%%%%%%%%%%%%%%%%%%%%%%%%%%%%%%%%%%%%%%%%%%%%%%%%%%
%  Number dimension plots
%%%%%%%%%%%%%%%%%%%%%%%%%%%%%%%%%%%%%%%%%%%%%%%%%%%%%%%%%%%%%%%%%%%%%

\newpage 

%%%%%%%%%%%%%%%%%%%%%%%%%%%%%%%%%%%%%%%%%%%%%%%%%%%%%%%%%%%%%%%%%%%%%

\newpage

%%%%%%%%%%%%%%%%%%%%%%%%%%%%%%%%%%%%%%%%%%%%%%%%%%%%%%%%%%%%%%%%%%%%%


\begin{references}
\bibitem{Text1} The precise definition of a $k$-block is(previous
reference): A graph $G$ is a $k$-block if it has no $n$-separation for
any $n<k$. An $n$-separation of $G$ is a pair of subgraphs of $G$, $H$,
and $K$, where (i) $H\cup K =G$, (ii) $H$ and $K$ have exactly $k$
common vertices, and (iii) $H$ and $K$ each have at least $k$ edges.
\end{references}
\end{document}